
\font\bif=cmbx12 at 14pt\pageno=0 \vbadness=10000

{\baselineskip=16pt \nopagenumbers \null \vfill
\centerline{\bif On Physical Equivalence between Nonlinear
Gravity Theories} \centerline{\bif and a General--Relativistic
Self--Gravitating Scalar Field} \vskip0pt plus0.5fill
\centerline{\bf Guido MAGNANO} \centerline{\bf Leszek M.
SOKO\L OWSKI\footnote{$^*$} {\rm On leave of absence
from:\hfill\break\indent Astronomical Observatory,
Jagellonian University, Orla 171, 30--244 Krakow, Poland}}
\centerline{Istituto di Fisica Matematica ``Joseph--Louis
Lagrange'', via Carlo Alberto 10, I--10123 Torino, Italy}\medskip
\vfill {\bf Abstract:} We argue that in a nonlinear gravity theory
(the Lagrangian being an arbitrary function of the curvature
scalar $R$), which according to well--known results is
dynamically equivalent to a self--gravitating scalar field in
General Relativity, the true physical variables are exactly those
which describe the equivalent general--relativistic model (these
variables are known as Einstein frame). Whenever such
variables cannot be defined, there are strong indications that the
original theory is unphysical, in the sense that Minkowski space
is unstable due to existence of  negative--energy solutions close
to it. To this aim we first clarify the global net of relationships
between the nonlinear gravity theories, scalar--tensor theories
and General Relativity, showing that in a sense these are
``canonically conjugated'' to each other. We stress that the
isomorphisms are in most cases local; in the regions where these
are defined, we explicitly show how to map, in the presence of
matter, the Jordan frame to the Einstein one and backwards. We
study energetics for asymptotically flat solutions for those
Lagrangians which admit conformal rescaling to Einstein frame
in the vicinity of flat space. This is based on the second--order
dynamics obtained, without changing the metric, by the use of a
Helmholtz Lagrangian. We prove for a large class of these
Lagrangians that the ADM energy is positive for solutions close
to flat space, and this is determined by the lowest--order terms,
$R+aR^2$, in the Lagrangian. The proof of this Positive Energy
Theorem relies on the existence of the Einstein frame, since in
the (Helmholtz--)Jordan frame the Dominant Energy Condition
does not hold and the field variables are unrelated to the total
energy of the system. This is why we regard the Jordan frame as
unphysical, while the Einstein frame is physical and reveals the
physical contents of the theory. The latter should hence be
viewed as {\it physically\/} equivalent to a self--interacting
general--relativistic scalar field. \bigskip \noindent{\bf PACS
No.: 04.50.+h}   \vskip0pt plus0.5fill\eject} \baselineskip=12pt
\newcount\fnc \fnc=1
\def\foot#1{\footnote{$^{\the\fnc}$}{#1}\advance\fnc by1}
\font\bfi=cmbxti10 \def\drulefill{\leaders \hbox{$\hbox to
2pt{\hrulefill}\llap{\raise3pt\hbox to 2pt{\hrulefill}}$}\hfill}
\def\en#1{\eqno{(#1)}} \parskip=5pt \def\exp(#1){e^{#1}}
\def\lagrangian{Lagrangian} \def\HL{Helmholtz \lagrangian}
\def\LH{L_{_H}} \def\NL{L_{_{\rm NL}}} \def\SNL{S_{_{\rm
NL}}} \def\lmat{\ell_{\rm mat}} \def\Lvac{L_{\rm vac}}
\def\Lphi{L_{\varphi}} \def\Lchi{L_{\chi}} \def\Lg{L_{g}}
\def\g{\tilde g} \def\hg{\hat g} \def\w{\tilde w}
\def\tu{\tilde u} \def\nrho{\tilde\rho} \def\G{\tilde G}
\def\L{\tilde L} \def\U{\tilde U} 
\def\R{\tilde R} \def\hR{\hat R}
 \def\nnabla{\tilde\nabla}
\def\half{{1\over2}}
\def\section#1{\bigskip\medskip\goodbreak
\par\noindent{\bf #1}\medskip}
\def\subsection#1{\medskip\goodbreak\par\noindent{\bfi
#1}\medskip}
\def\dal{{\offinterlineskip\lower1pt\hbox{\kern2pt\vrule
width0.8pt \vbox to8pt{\hbox to6pt {\leaders\hrule
height0.8pt\hfill}\vfill \hbox to6pt{\hrulefill}}\vrule
\kern3pt}}} \def\ndal{\tilde\dal} \def\sqg{\sqrt{-g}\,}
\def\nsqg{\sqrt{-\g}\,} \def\hsqg{\sqrt{-\hg}\,}
\def\sqf{{\scriptstyle \sqrt{2\over3}}}
\def\vder#1#2{{\delta#1\over\delta#2}}

\def\tder#1#2{{d#1\over d#2}}
\def\pder#1#2{{\partial#1\over\partial#2}}
\def\su#1{\vskip-18pt\noindent #1\vskip-18pt}
\def\ssu#1{\vskip-12pt\noindent #1\vskip-12pt}
\def\etal{{\it et al.\/}} \def\sth{\sqrt{\scriptstyle{3\over 2}}}
\def\stt{\sqrt{\scriptstyle{2\over 3}}}

\section{1. Introduction}

Metric theories of gravitation which are based on a
\lagrangian\ density depending in a nonlinear way on the
scalar curvature -- such theories, somehow improperly, are
usually called ``nonlinear gravity" (NLG) models -- share a
general property, which has been extensively described in
several works: acting on the metric by a suitable conformal
transformation, the field equations can be recast into Einstein
ones for the rescaled metric, interacting with a scalar field. It is
therefore claimed that any NLG theory is equivalent to General
Relativity (with the scalar field).

A similar phenomenon occurs in Jordan--Brans--Dicke theory
and its generalizations, the scalar--tensor gravity theories (STG).
The original pair of variables (metric + scalar field) can be
replaced by a new pair in which the metric has been conformally
rescaled, and in the new variables the field equations become
those of General Relativity: the scalar field, which is not affected
by the transformation, turns out to be minimally coupled to the
rescaled metric. The original set of variables is commonly called
{\it Jordan conformal frame}, while the transformed set, whose
dynamics is described by Einstein equations, is called {\it
Einstein conformal frame}. A problem thus arises, whether the
tensor representing the {\it physical} metric structure of
space--time is the one belonging to the Jordan frame or to the
Einstein frame.

This problem can be traced back to Wolfgang Pauli in the early
fifties (quoted in [1]). In a system consisting of metric gravity
and a scalar field there is an ambiguity: the metric tensor can be
conformally rescaled by an arbitrary (positive) function of the
scalar. Thus, besides the original (Jordan) and  Einstein frames,
there exists an infinite number of conformally--related frames,
each consisting of a pair $(F(\varphi)g_{\mu\nu},\varphi)$
with a different $F$. One asks: are the metrics in the frames
$(g_{\mu\nu},\varphi)$ and
$(F(\varphi)g_{\mu\nu},\varphi)$ also {\it physically\/}
equivalent? The same question would clearly arise also for a
more general change of variables
$g'_{\mu\nu}=g'_{\mu\nu}(g_{\alpha\beta},\varphi)$,
$\varphi'=\varphi'(g_{\alpha\beta},\varphi)$.

NLG and STG theories are in fact deeply connected, as is
described in this paper, and can actually be viewed as different
versions of the same model. We now define the notion of Jordan
frame and  Einstein frame for NLG theories; the main purpose of
this paper is to analyze the problem of determining which
``frame"\foot{We use here the word ``frame" to denote a choice
of dynamical variables, rather than a choice of a reference frame
in space--time; this might seem misleading, however this
terminology is commonly used in the previous literature on the
subject. Sometimes the term ``gauge'' (also abused in this
context) is used instead of ``frame'' [10].}\ is the {\it physical}
one. By ``physical frame'' we mean a set of field variables which
are (at least in principle) measurable and satisfy all general
requirements of classical field theory, e.g.~give rise to positive--
definite energy density (we are aware of the fact that the term
``physical'' is frequently abused in the literature).

We assume, for simplicity, that the space--time is four--
dimensional, although all the calculations can be actually carried
out in higher dimension without significant modifications. The
signature is $( -+++)$ and we set $\hbar=c=8\pi G=1$. Let us
consider the \lagrangian\ of a {\it vacuum} NLG theory,   $$
L=f(R)\sqg ,\en{1.1}  $$  which generates the fourth-order
equations $$ f'(R)R_{\mu\nu}-\half f(R)g_{\mu\nu}-
\nabla_{\mu}\nabla_{\nu}f'(R) +g_{\mu\nu}\dal f'(R)=0 ,
\en{1.2} $$ where $f'(R)\equiv {df\over dR}$ and $\dal\equiv
g^{\mu\nu}\nabla_{\mu}\nabla_{\nu}$. In the vacuum case,
the Jordan frame includes only the metric tensor $g_{\mu\nu}$.
According to a well--known procedure [2--4], we introduce a
pair of new variables $(\g_{\mu\nu}, p)$, related to
$g_{\mu\nu}$ (and to its derivatives) by  $$ p = f'(R)
,\qquad\qquad \g_{\mu\nu} = p g_{\mu\nu} .\en{1.3} $$
(this transformation was rediscovered several times and
generalized to the case of \lagrangian s depending also on
$\dal^k R$, [5, 6]) The scalar $p$ is dimensionless, and to
ensure the regularity of the conformal rescaling  it is usually
assumed that $p>0$ (we will retake this point below). Then the
two metrics $g_{\mu\nu}$ and $\g_{\mu\nu}$ have the same
signature.  Let $r(p)$ be a solution of the equation $f'[r(p)]-p=0$
(it may not be unique, and we will consider this problem later);
the field equations for the new variables become  the following
(second-order) ones: $$ \G_{\mu\nu} = p^{-
2}\left\lbrace{3\over2}(p_{,\mu}p_{,\nu}-
\half\g_{\mu\nu}\g^{\alpha\beta}p_{,\alpha}p_{,\beta})  +
\half \Big(f[r(p)]-p\cdot r(p)\Big)\g_{\mu\nu}\right\rbrace
,\en{1.4a}   $$ whereby $\G$ is the Einstein tensor of the
rescaled metric $\g_{\mu\nu}$, and $$ \ndal p - p^{-
1}\left\lbrace\g^{\mu\nu}p_{,\mu}p_{,\nu} + {1\over
3}\left(2f[r(p)]- pr(p)\right)\right\rbrace = 0 . \en{1.4b}  $$
These equations can be derived from the \lagrangian\ [3] $$
\L=\R\nsqg-\left\lbrace{3\over 2}p^{-
2}\g^{\mu\nu}p_{,\mu}p_{,\nu} +p^{-1} r(p)-p^{-
2}f[r(p)]\right\rbrace\nsqg .\en{1.5}  $$ It is customary to
redefine the scalar field by setting $p=\exp(\sqf\phi)$; the
\lagrangian\ then becomes  $$ \L=\left[\R-
\g^{\mu\nu}\phi_{,\mu}\phi_{,\nu}-2V(\phi)\right]\nsqg
;\en{1.6}  $$ one sees that $\phi$ is minimally coupled to
$\g_{\mu\nu}$ and is an ordinary massive self--interacting
scalar field with a potential  $$ V(\phi)=\half\exp(-
\sqf\phi)\,r[p(\phi)]-\half\exp(-2\sqf\phi)f(r[p(\phi)])\ ,
\en{1.7} $$  which is determined by the original \lagrangian\
(1.1). The variables $(\g_{\mu\nu}, \phi)$ provide the Einstein
frame for the NLG theory.

In the Jordan frame, gravity is entirely described by the metric
tensor $g_{\mu\nu}$. In the Einstein frame, the scalar field
$\phi$ acts as a source for the metric tensor $\g_{\mu\nu}$
and formally plays the role of an external ``matterfield";
however, the original theory did not include any matter, hence
we are led to regard the scalar field occurring in the Einstein
frame (which corresponds to the additional degree of freedom
due to the higher order of the field equations in  Jordan frame)
as a ``non--metric" aspect of the gravitational interaction. From
this viewpoint, the NLG theory, although {\it mathematically}
equivalent to  General Relativity, is physically different because
in the Einstein frame, where the equations coincide with those of
GR, gravity is no longer represented by the metric tensor alone.
However, another viewpoint is possible: one could assume that
the fourth-order picture in Jordan frame represents an "already
unified" model including a non-gravitational degree of freedom
(a minimally coupled nonlinear scalar field), and that the
gravitational interaction is described only by the rescaled metric
in the Einstein frame (see Appendix D).

This clearly indicates that the problem of the physical nature of
the fields occurring in both frames should be addressed {\it
prior} to any other consideration on the physical significance of
the equivalence between NLG and GR. From the outline
presented above, one might be led to the following conclusion: if
one starts from a {\it vacuum} NLG theory, there is no way to
decide {\it a priori} which frame should be taken as the
``physical" one; the choice of the physical metric is an additional
datum, which affects the physical interpretation of the model
but is essentially independent of the formal structure of the
theory. On the other hand, if we formulate a gravitational theory
including from the very beginning matter fields, the ambiguity
is broken by the coupling of the metric tensor with such matter
fields. In fact, the two metric tensors, $g_{\mu\nu}$ and
$\g_{\mu\nu}$, interact in a different way with each external
field (we will discuss this point in greater detail below),
therefore one should be able to single out the physical metric by
requiring that matter fields be minimally coupled with it, and
that neutral test particles fall along its geodesic lines.

This viewpoint, more or less explicitly formulated, seems to be
shared by many authors. However, a criterion based on the
coupling of matter with gravity would be effective only if the
total \lagrangian\ were determined on independent grounds,
e.g.~by string theory. As a matter of fact, this is not so: the
theories which we consider in this paper are commonly viewed
as fundamental ones and constructed by adding usual
interaction \lagrangian s to a purely gravitational one. The
point, that some authors seem to overlook, is that adding
minimal--coupling terms to the \lagrangian\ (1.1) already
entails that the Jordan frame is assumed to be the physical one.
In this situation, claiming  that the Einstein frame is unphysical,
because the coupling between matter and the rescaled metric is
not the usual one, is a {\it petitio principii\/}. It would be
equally reasonable to add standard interaction terms, minimally
coupled to  $\g_{\mu\nu}$, to the \lagrangian\ (1.6): the
resulting coupling with $g_{\mu\nu}$ would then turn out to
be unphysical. In other terms, the frequently repeated argument
[7] in favour of the Jordan frame, namely that physical
measurements are made in this frame (``atomic frame'') since
atomic masses are physical constants there, can be equally well
used in favour of the Einstein frame, because the argument is a
direct consequence of the arbitrary (at this level of reasoning)
choice of the full action for a gravitational theory. Thus, the
ambiguity is still present.

Indeed, this ambiguity is faithfully represented in the literature.
The authors dealing with nonlinear or scalar--tensor theories of
gravity in the context of cosmology or of high--energy physics
can be broadly divided into four groups. According to the
authors of the first group, the Jordan frame is the physical one
and the Einstein frame merely serves as a practical
computational tool\foot{However, Kalara {\it et al.\/} [7],
while investigating the power--law inflation in the Einstein
frame, interpret the solutions and fit their parameters as if this
frame were the physical one.}[7, 8, 9]. (Barrow and Maeda [7]
remark that the Einstein frame is computationally advantageous
only in a vacuum theory. If matter, e.g.~a perfect fluid, is
included, the conformal transformation usually does not lead to
simpler equations since in the Einstein frame the scalar field is
coupled to the fluid). The authors in the second group regard
the Einstein frame as being physical, either because of its
resemblance to General Relativity\foot{Gibbons and Maeda [10]
admit that there might be an argument (or merely a someone's
prejudice) stating that the ``physically correct'' frame is different
from the Einstein one.}[10, 11], or since the standard formalism
for quantizing the scalar field fluctuations in the linear
approximation does not apply in the Jordan frame (or at least is
suspect there) [12, 13], or because the massless spin--two
graviton in Jordan--Brans--Dicke theory is described by the
Einstein--frame metric [14], or finally because this is implied by
dimensional reduction of a higher--dimensional action [15]. The
third group consists of the authors claiming that the two frames
{\it are\/} physically equivalent, at least at the classical level,
since conformal transformations do not change the mass ratios
of elementary particles and therefore does not alter
physics\foot{Garay and Garcia--Bellido in [16] introduce a
concept of ``physical frame'' which is different from ours.
According to them the ``physical'' frame is one ``in which
observable particles have constant masses, since in this frame
particles follow geodesics of the metric''. Then, by the
assumption on the form of the full action, the Jordan frame
coincides with the ``physical'' one. We notice at this point that in
General Relativity particles (with constant masses) usually do
not move along geodesics: consider for instance a perfect fluid
with pressure. Geodesic motion is rather exceptional, it occurs in
the case of pressureless dust and for non--interacting test
particles. }(``physics cannot distinguish between conformal
frames'') [16]. The last, rather inhomogeneous group, involves
authors who either use both conformal frames without
addressing the problem of which of them (if any) is physical, or
work exclusively in the original Jordan frame without making
any reference to the existence of the rescaled metric [17].

In our opinion, the strongest argument which exists in the
literature in favour of one of the conformal frames (if the
universe is exactly four--dimensional) is the one based on
quantization of field fluctuations. ``It appears as if the
quantization and conformal transformation are two mutually
noncommutable procedures'' [13]. Here we show that classical
relativistic field theory provides another argument which points
to the same conformal frame.

The main arguments presented in this paper can be outlined in
the following statements: \item{(i)} In contrast with claims
raised in the previous literature, nonlinear and scalar--tensor
theories of gravity can be equally well formulated, as far as {\it
formal consistency\/} is considered, in terms of either one of the
frames; from this viewpoint, either frame can in principle be
assumed to be the physical one; \item{(ii)} However, the physics
described by nonlinear and scalar--tensor theories of gravity is
not conformally invariant and using different conformal frames
one finds inequivalent effects. Therefore, choosing a correct
(i.e.~physical) frame is an indispensable part of theoretical
investigation [18, 19]; \item{(iii)} The physical metric should be
singled out already in the vacuum theory; the coupling of a
given metric to matter fields is in fact {\it determined\/} by the
physical significance ascribed to it, i.e.~by its relation to the
physical metric.  \item{(iv)} In particular, if the physical metric
is assumed to be the rescaled (Einstein--frame) one, the original
nonlinear \lagrangian\ should include interaction with
ordinary matter in such a way that the corresponding coupling
with the Einstein--frame metric turns out to be minimal. This
problem has never been addressed in the previous literature. We
provide here a systematic and unique way to obtain \lagrangian
s which reduce to any given vacuum nonlinear \lagrangian\ in
the absence of matter, and reproduce upon a suitable rescaling
any (with some restrictions) minimally coupled matter
\lagrangian; \item{(v)} Since consistency arguments do not
allow one to establish which set of variables is physical, we
investigate a different criterion. This criterion is purely classical
and is provided by the positivity of energy for small fluctuations
of {\it every} dynamical variable around the ground state
solution;  \item{(vi)} The notion of gravitational energy in NLG
theories is inapplicable to this aim, due to the lack of a positivity
theorem for higher--order gravitational theories. To circumvent
this problem, one should compare two {\it second--order\/}
versions of the same theory, namely the second--order dynamics
obtained in the Einstein frame and the ``Hamiltonian''
formulation of the theory constructed -- {\it without\/}
rescaling the metric -- through a Legendre transformation;
\item{(vii)} We show that the Positive Energy Theorem for
Nonlinear Gravity, proven by Strominger [20] for a quadratic
case, holds for a larger class of \lagrangian s for which the
Einstein frame can be defined around flat space. Actually,
whether the ADM energy is positive or not depends on the
potential (1.7). The existence of the Einstein frame is in any case
essential for assessing classical stability of Minkowski space and
positivity of energy for nearby solutions. In the Jordan frame,
the Dominant Energy Condition never holds. For these reasons,
the Einstein frame is the most natural candidate for the role of
physical frame in NLG theories.

The paper is organized as follows. In Section 2, we discuss in
full detail the statements (i), (ii) and (iii) above, using examples
taken from the literature. Some material is alrady known,
nevetheless the amount of confusion existing in the recent
literature on the subject justifies a detailed exposition. We
provide there a global and complete picture of relationships
between NGL and STG theories and general--relativistic models
of a scalar field, while previously only separate connections
were known. Section 3 is the heart of the paper. We discuss
there items (v), (vi) and (vii). We investigate there a large class
of nonlinear \lagrangian s, for which flat space (in the Jordan
frame) is a solution and the Einstein frame exists for it. While the
ADM energy can be defined for any asymptotically flat solution,
one is unable to establish its sign in the Jordan frame. In the
Einstein frame, the relation betwen the interior of the system
and its total energy takes on the standard form known from
General Relativity. This fact convinces us that the Jordan frame
is unphysical. Section 4 contains conclusions. All technical parts
of the paper are moved to appendices and the main body of the
paper can be read without consulting them In particular,
Appendix C provides theoretical background for some results
applied in Sections 2 and 3.  The ``inverse problem of nonlinear
gravity'' (finding the purely metric nonlinear \lagrangian\
which generates a prescribed potential for the scalar field) is
presented in Appendix D, and the difficulties arising from
possible local failures of the conditions ensuring the existence of
the Einstein frame (a problem sometimes raised in the literature,
but still lacking a systematic investigation) are dealt with in
Appendix E.

\vfill\eject

\section{2. Interaction of gravity with matter and conservation
laws}

Some authors have tried to solve the problem of determining the
physical metric by showing that only one possible choice allows
one to obtain a divergenceless energy--momentum tensor for
any self--gravitating matter. Brans used this argument to claim
that only the Jordan frame is physical [18], while in a recent
paper [21] Cotsakis was led by a more detailed investigation of
this point to the conclusion that the Eistein frame is the physical
one. Unfortunately, an error invalidates Cotsakis'
proof\foot{The crucial step of the proof consists in taking the
divergence of the fourth--order equation (eq. (2.2) below):
Cotsakis claims that it does not vanish in general. Actually, it is
a generalized Bianchi identity (see Appendix A).}. Here we
show that studying the conservation laws for matter does not
allow one to find out which frame is physical: the equations of
motion for matter and gravity form consistent and closed
systems for both Jordan and Einstein frames.

Our basic assumption is that {\it the gravitational interaction of
matter should bedescribed by minimal coupling with the {\rm
physical} metric tensorfield}. In other words, the physical metric
of space--time should be identified prior to the construction of
the \lagrangian\ for a gravitating system. Different
identifications of the physical frame will give rise to
(mathematically and physically) inequivalent \lagrangian s.
This assumption is not commonly accepted; e.g., the authors of
[13] argue that the Einstein frame is physical, but nevertheless
they assume that matter minimally couples to the metric in the
(unphysical) Jordan frame. A similar viewpoint is taken by
Alonso \etal\ in [11].

The assumption is based on the postulate that the great
advantage of Einstein's General Relativity, i.e.~the universal
validity of the minimal coupling principle, should be retained in
NLG and scalar--tensor theories of gravity. The principle is
partially abrogated in the so--called ``extended Jordan--Brans--
Dicke theory'', where the ordinary (``visible'') and dark
(``invisible'') matter minimally couple to different (conformally
related) metric fields (Damour \etal\ in [7]). Therefore,
although observationally viable, we do not take this theory into
account in this paper, on theoretical grounds.

In the first part of this Section, we discuss the implication of our
assumption in the context of NLG theory, considering separately
the two cases: either the Jordan--frame or the Einstein--frame
metric is regarded as physical. We show that the two cases
describe different physical models of gravitational interaction of
matter fields, and we show that {\it each\/} of the two models
can be consistently formulated in {\it both\/} frames.

The scalar--tensor gravity (STG) theories share with NLG
theories the feature that they can be reformulated as a general-
relativistic model for a self--gravitating scalar field\foot{It is
worth noticing that we should discriminate between the
dynamical equivalence of \lagrangian s and equality of the
action integrals, see Appendices C and E. This difference is
significant in quantum theory; in classical field theory it is
irrelevant as physical meaning is given only to solutions of the
field equations.}. As a matter of fact, any NLG theory can be also
in terms of a scalar--tensor theory with nontrivial cosmological
function (without changing the metric). This is explained in
detail in Appendix C, and is extensively used in Section 3. Few
authors seem to be aware of this relation, while many authors
establish a connection between NLG and STG theories via the
equivalence to General Relativity. Hence, NLG and STG theories
appear strictly intertwined in the literature, sometimes causing
confusion. A physical difference between STG and NLG theories
is that in the former it is {\it a priori\/} postulated that matter
couples minimally to the Jordan--frame metric. The dynamical
equivalence clearly shows that, despite appearances, there is no
deeper difference between the two classes of gravity theories.
Nevertheless, for clarity reasons, we also discuss here
interactions with matter and conservation laws for STG theories.

As it is shown in Appendix D, the ``inverse problem of
nonlinear gravity'', i.e.~finding a NLG \lagrangian\ which is
equivalent to General Relativity plus a scalar field with any self-
-interaction potential $\U(\phi)$, has a solution. An exception is
provided by the massless linear scalar field: if $\U(\phi)=0$ and
the cosmological constant vanishes, the equivalent nonlinear
\lagrangian\ does not exist and the scalar field cannot be
absorbed into the Jordan--frame metric. However, the massless
scalar field is conformally equivalent to both a STG theory and
to the conformally--invariant scalar field model.

First we investigate the consequences of our assumption for the
case of NLG theory.

\subsection{2.1 NLG theory, case I: the original metric
$g_{\mu\nu}$ is physical.}

\noindent According to the assumption, the \lagrangian\ for
gravity and matterfields (collectively denoted by $\Psi$)
reads\foot{The coefficient in front of $\lmat$ is chosen on the
assumption that the linear term in the Taylor expansion of
$f(R)$ has coefficient 1 (see Appendix A).} $$
L=[f(R)+2\lmat(\Psi, g)]\sqg .\en{2.1}  $$ The gravitational
field equations are then $$ \vder{}{g^{\mu\nu}}[\sqg
f(R)]=f'(R)R_{\mu\nu}-\half f(R)g_{\mu\nu}-
\nabla_{\mu}\nabla_{\nu}f'(R) +g_{\mu\nu}\dal
f'(R)=T_{\mu\nu}(\Psi, g)\ , \en{2.2} $$ where, as usual,
$T_{\mu\nu}=-
{2\over\sqg}\vder{}{g^{\mu\nu}}(\sqg\lmat)$. It follows from
Noether theorem (which in this case is equivalent to a
generalized Bianchi identity) that
$\nabla_{\nu}T^{\mu\nu}=0$ (see Appendix A).  Using the
general procedure described in Appendix C and upon
conformal rescaling (1.3) of the metric one gets the Einstein--
frame \lagrangian\ for the system (metric + scalar field +
matter) [3] $$ \L(\g,\phi,\Psi)=\left[\R(\g) -
\g^{\mu\nu}\phi_{,\mu}\phi_{,\nu} -2V(\phi) + 2\exp(-
2\stt\phi)\lmat(\Psi,\exp(-\stt\phi)\g)\right]\sqg ,\en{2.3}
$$ which is dynamically equivalent to the \lagrangian\ (2.1).
Here, as before, $\phi\equiv\sth\ln p$ is a scalar field having
the canonical kinetic term and self--interacting via the potential
$V(\phi)$ given in (1.7). The field equations for the metric and
the scalar in the Einstein frame can be obtained either by
transforming equations (2.2) or directly from the \lagrangian\
(2.3). In the latter case, caution is needed whilst taking variations
of the matter part, to avoid ambiguities in the definition of the
stress tensor for the matter field $\Psi$. In fact, defining
separate stress tensors for $\phi$ and for $\Psi$ makes sense
only when the matter \lagrangian\ can be split in two parts,
each one depending only on the metric and on one of the fields.
Therefore, it is advisable to formulate the gravitational
equations in both frames in terms of the tensor $T_{\mu\nu}$
already defined in terms of the physical metric\foot{We notice
that the definition provides the covariant components of the
stress tensor; if contravariant components $T^{\mu\nu}$ are
defined by raising the indices with the physical metric
$g^{\mu\nu}$, then the contravariant form of  equation (2.4)
reads \def\stt{\gamma} $ \G^{\mu\nu} =
t^{\mu\nu}(\phi,\g) + \exp(-3\stt\phi)
T^{\mu\nu}(\Psi,\exp(-\stt\phi)\g) . $}. To this aim, one
should first take the variation of the matter term with respect to
$\Psi$ and to $g^{\mu\nu}$ and then use $\delta
g^{\mu\nu}=p\delta\g^{\mu\nu}+ \g^{\mu\nu}\delta p$.
As a result, one finds the field equations for the metric in the
form $$ \G_{\mu\nu} = t_{\mu\nu}(\phi,\g) + \exp(-
\stt\phi) T_{\mu\nu}(\Psi,\exp(-\stt\phi)\g) ,\en{2.4}   $$
where $\stt=\sqrt{2\over3}$ and $$ t_{\mu\nu} =
\phi_{,\mu}\phi_{,\nu}-
\half\g_{\mu\nu}\g^{\alpha\beta}\phi_{,\alpha}
\phi_{,\beta} - V(\phi)\g_{\mu\nu} \en{2.5} $$ plays the role
of the effective stress-energy tensor for the scalar $\phi$, and
the equation of motion for $\phi$ $$ \ndal\phi =
\tder{V}{\phi} + {1\over{\scriptstyle \sqrt{6}}}\exp(-
\stt\phi)T_{\mu\nu}(\Psi,\exp(-\stt\phi)\g)\ ,\en{2.6} $$
\su{with} $$ \qquad\tder{V}{\phi}={1\over{\scriptstyle
\sqrt{6}}p}\left[{2\over p}f[r(p)]-r(p) \right] \qquad\quad
\left(p=p(\phi)=\exp(\stt\phi)\right)\ . \en{2.7} $$ The
Bianchi identity now implies  $$
\nnabla^{\nu}\left[t_{\mu\nu}(\phi,\g) + \exp(-\stt\phi)
T_{\mu\nu}(\Psi,\exp(-\stt\phi)\g)\right]=0 , \en{2.8} $$ and
the two stress tensors are not separately conserved, since the
scalar field interacts with matter, as is explicitly seen in
\lagrangian (2.3).

In this picture the scalar field $\phi$ influences the motion of
any gravitating matter, except for the particular case in which
the interaction lagrangian $\lmat$ is conformally invariant (see
Appendix B). From the {\it physical\/} viewpoint this theory is
{\it not\/} equivalent to General Relativity: in fact, gravity is
completely represented by a metric tensor only in the Jordan
frame (where it obeys fourth-order equations), while in the
Einstein frame, where equations of motion for $\g_{\mu\nu}$
are formally Einstein ones, there is a non--geometric
gravitational degree of freedom, represented by the scalar
$\phi$, which is universally coupled to matter. From the
viewpoint of conservation laws, however, no inconsistency
arises from the assumption that the original metric
$g_{\mu\nu}$ be the physical one.

\subsection{2.2 NGL theory, case II: the conformally rescaled
metric $\g_{\mu\nu}$ is physical.}

\noindent The original metric $g_{\mu\nu}$ plays now the role
of a variable providing an ``already unified" fourth--order
version of a theory including the gravitational metric plus a
nonlinear scalar field (these theories form a basis for inflationary
cosmological models). The vacuum \lagrangian\ (1.1) should
first be transformed into the corresponding Einstein--frame
\lagrangian\ (1.6), and {\it then\/} one has to add the
interaction lagrangian for matter, minimally coupled to
$\g_{\mu\nu}$: $$ \L=\left[\R -
\g^{\mu\nu}\phi_{,\mu}\phi_{,\nu} -2V(\phi) +
2\lmat(\Psi,\g)\right]\nsqg\ .\en{2.9}  $$ The matter
\lagrangian\ $\lmat$ is $\phi$--independent since there is no
physical motivation to assume that matter interacts with the
scalar. Let us stress that in both cases $\lmat$ is constructed in
the same way: first one establishes by physical considerations
the form of the free \lagrangian\ (or of the interaction
\lagrangian\ for a number of coupled fields, e.g.~for a complex
scalar field minimally coupled to electromagnetism) for $\Psi$
in Minkowski space and then one replaces the flat metric by the
metric which is viewed as physical. Therefore $\lmat(\Psi,g)$ in
Case I and $\lmat(\Psi,\g)$ of Case II are the same fuctions of
their respective arguments; clearly, for conformally--related
metrics, $\lmat(\Psi,g)\ne\lmat(\Psi,\g)$.

The field equations resulting from (2.9) now read $$\eqalignno{
\G_{\mu\nu} &= t_{\mu\nu}(\phi,\g) + T_{\mu\nu}(\Psi,\g)
&(2.10)\cr \ndal\phi &= \tder{V}{\phi}&(2.11)\cr
\vder{}{\Psi}(\lmat\nsqg) &=0 & (2.12)}   $$ where
$t_{\mu\nu}$ is given in (2.5), with the potential as in (1.7) and
(2.7).

Here $T_{\mu\nu}\equiv -{2\over \nsqg
}\vder{}{\g^{\mu\nu}}(\lmat\sqg)$, and due to the absence of
any interaction between $\phi$ and $\Psi$ not only the total
stress--energy tensor $t_{\mu\nu} + T_{\mu\nu}$ is conserved,
but the stress tensors of each of the fields are separately
conserved,
$\nnabla^{\nu}t_{\mu\nu}=\nnabla^{\nu}T_{\mu\nu}=0$.

To find the interaction of matter with the original metric
$g_{\mu\nu}$ one might naively trasform back the equation
(2.10) to the Jordan frame, eliminating the scalar field $\phi$ by
the relations $\exp(\stt\phi)=p=f'(R)$ and $r[p(\phi)]=R(g)$;
one would get $$ f'(R)R_{\mu\nu}-\half f(R)g_{\mu\nu}-
\nabla_{\mu}\nabla_{\nu}f'(R) +g_{\mu\nu}\dal f'(R) =
f'(R)T_{\mu\nu}(\Psi,f'(R)g) , \en{2.13} $$ where
$T_{\mu\nu}$ is the matter stress tensor defined in the Einstein
frame and expressed in terms of the original unphysical metric
$g_{\mu\nu}$. In general,  however, equation (2.13) is {\it
incorrect\/}. To see this, let us first recast the field equations
(2.10) and (2.11) in terms of the variables $(p,g_{\mu\nu})$, {\it
without\/} assuming that $r(p)=R(g)$. These read
$$\eqalignno{ G_{\mu\nu}(g) &= p^{-
1}(\nabla_{\mu}\nabla_{\nu}p - g_{\mu\nu}\dal p) -
pV(p)g_{\mu\nu} + T_{\mu\nu}(\Psi, pg)\ , &(2.14)\cr \dal p
&= {2p^3\over 3} \tder{V}{p}\ .&(2.15)}   $$ Taking the trace of
(2.14), eliminating $\dal p$ with the aid of (2.15) and using the
explicit form of the potential, $V={1\over 2p}r(p)-{1\over
2p^2}f[r(p)]$, one arrives at the equation  $$ R(g)-
r(p)+g^{\mu\nu}T_{\mu\nu}(\Psi,pg)=0\ . \en{2.16} $$ Recall
that $r(p)$ is defined by solving the equation $f'[r(p)]=p$. In the
absence of matter or if its stress tensor is traceless, (2.16) yields
$R(g)=r(p)$, then $p=f'(R)$ and equations (2.13) do hold. In
general, however, setting $p=f'(R)$ is inconsistent with (2.16).
Provided that $T_{\mu\nu}$ does not contain covariant
derivatives of $\Psi$, (2.16) can be viewed as an {\it
algebraic\/} equation for the scalar field $p$. Solving this
equation for $p$ provides, in the presence of matter, the correct
relation $p=P(R;g,\Psi)$ which allows one to re--express the
scalar field $p$ in terms of the curvature scalar $R(g)$ and
obtain higher--order equations of motion in terms of $g$ and
$\Psi$ only.  To avoid confusion, let us call from now on the
original unphysical frame, in which the vacuum \lagrangian\
takes the form $L=f(R)\sqg$, the {\it vacuum Jordan conformal
frame\/} (VJCF), and the (also unphysical) frame into which the
scalar field $\phi$ can be re--absorbed, in the presence of the
matter term in (2.9), the {\it matter Jordan conformal frame\/}
(MJCF). Except for conformally invariant material systems, the
two Jordan frames are different. A deeper understanding of the
reason why the two frames do not coincide is provided by the
Legendre--transformation method. The explicit construction of
the MJCF, the corresponding nonlinear \lagrangian\ and the
resulting field equations are given in Appendix C. In MJCF,
being the ``already unified'' frame, matter is nonminimally
coupled to the metric and it is a generic feature that in the
absence of gravitation, in flat spacetime, the nonlinear
\lagrangian\ does not reduce to the standard form for a given
species of matter. This fact significantly influences conservation
laws. As is shown in Appendix C, it is possible to separate a
``purely gravitational part'' in the gravitational field equations in
MJCF; this part satisfies the generalized Bianchi identity. Four
matter conservation laws then follow from the equations in the
same way as in General Relativity. These involve a number of
terms mixing the curvature scalar with the matter variables and
consequently do not resemble at all the elegant conservation
laws $T_{\mu\nu}{}^{;\nu}=0$ of Einstein's theory. One can
only learn from them that in this frame the matter worldlines
explicitly depend on curvature except for massless particles
(photons), see Appendix B.

In the Einstein frame, on the other hand, the Strong Equivalence
Principle holds; the only possible difference between this
version of gravity and General Relativity may arise from the
physical interpretation given to the scalar field, whose entire
role is confined to influence the metric field. In fact, $\phi$ is
assumed to describe a non--geometric spin--zero component of
gravitation [22, 23].

\subsection{2.3 Frames and conservation laws in STG theories}

Next we proceed to scalar--tensor theories of gravity (see [1, 24]
for recent reviews). These are conceptually different from NLG
theories because these are not purely metric gravitational
models, as the gravitational field is a doublet consisting of a
spin--two field and a (non--geometric) spin--zero field, the
Brans--Dicke scalar.

The action of a generic STG theory is (we use conventions of
Will's book [25]) $$ S=\int d^4x\sqg\left\lbrace{1\over
16\pi}\left[\varphi R-
{\omega(\varphi)\over\varphi}g^{\mu\nu}\varphi_{,\mu}
\varphi_{,\nu}
+2\varphi\lambda(\varphi)\right]+\lmat(\Psi,g)
\right\rbrace\ ;\en{2.17} $$ The ``cosmological function''
$\lambda(\varphi)$ is often omitted, and accordingly we
consider here the models in which $\lambda(\varphi)\equiv0$.
If the coupling function $\omega(\varphi)$ is constant, the
action is that of the Thiry--Jordan--Fierz--Brans--Dicke theory
(see [1] and [25] for references). The field variables
$g_{\mu\nu}$ and $\varphi$ form the Jordan conformal frame
(it is here that the concept originated). By assumption, ordinary
matter minimally couples to the metric $g_{\mu\nu}$ (and does
not couple to the scalar gravity), thus by this assumption the
Jordan frame is physical, i.e.~describes measurable spacetime
intervals. Proceeding as in the case of NLG theories one
introduces a scalar variable $$
p={1\over\sqg}\pder{}{R}L(g,\varphi,\Psi)= {1\over
16\pi}\varphi\ ;\en{2.18} $$ in this case it is not a function of
the curvature but it coincides (up to a constant factor) with the
already existing scalar field. With the aid of $p$ one defines a
new conformally--related metric $\g_{\mu\nu} = p
g_{\mu\nu}$, the transformation being already known since
1962 as {\it Dicke transformation\/} [26]. In terms of the new
variables $(\g_{\mu\nu},p)$ the action is (up to a full
divergence term) $$  S=\int d^4x\nsqg\left[\R-
\left(\omega(\phi)+{3\over 2}\right)p^{-
2}\g^{\mu\nu}p_{,\mu}p_{,\nu}+p^{-2}\lmat(\Psi,p^{-
1}\g_{\mu\nu})\right] \ ;\en{2.19} $$ and after a redefinition
of the Brans--Dicke scalar, $$ d\phi \equiv
\left(\omega(\phi)+{3\over
2}\right)^{1\over2}{d\varphi\over\varphi}\ ,
\qquad\omega>-{3\over 2} \en{2.20} $$ it takes the standard
form of the action for the linear massless scalar field minimally
coupled to the metric $$  S=\int d^4x\nsqg\left[\R-
\g^{\mu\nu}\phi_{,\mu}\phi_{,\nu}+\left({16\pi\over
\varphi(\phi)}
\right)^2\lmat(\Psi,{16\pi\over\varphi(\phi)}\g_{\mu\nu})
\right] \ .\en{2.21} $$ In terms of the Einstein--frame variables
$(\g_{\mu\nu},\phi)$ the matter part of the action describes an
interaction between ordinary matter and the scalar gravity.
Clearly, it is due to the use of the unphysical (by assumption)
variables for gravity. As a consequence the variational (with
respect to $\g_{\mu\nu}$) ``energy--momentum'' tensor for
$\Psi$ that can be defined by the \lagrangian\  $\varphi^{-
2}\lmat(\Psi,\varphi^{-1}\g_{\mu\nu})$ is different from the
matter stress tensor defined in the Jordan frame and
transformed to the Einstein frame; as already mentioned this
notion is rather ambiguous and of little use.

To study conservation laws for a STG theory one should first
correctly identify the energy--momentum tensor for the spin--
zero gravity. According to general rules of relativistic field
theory it is provided by the variational derivative of the
appropriate term in the \lagrangian. The latter should be
identified with care since the interaction is nonminimal. First,
the purely metric gravitational \lagrangian\ $\Lg$ should be
separated out. To this end one formally views the Brans--Dicke
scalar as a test field in a given fixed background $g_{\mu\nu}$.
The basic assumption of scalar--tensor gravity theories is that
the average value of the field $\varphi$ determines the present
value of the gravitational constant, $<\varphi>={1\over G}$.
Therefore, the situation where the scalar gravity is ``switched
off'' does not correspond to $\varphi\equiv 0$, but rather to
assuming that $\varphi$ is constant and equal to the present
value of ${1\over G}$ ($=8\pi$ in our units); notice that
$\varphi= 8\pi$ is actually a solution of the field equation only
if $R=0$. Setting $\varphi\equiv 8\pi$ and $\lmat\equiv 0$
reduces the \lagrangian\ in (2.17) to its purely metric part,
$\Lg=\half R\sqg $. This is the Einstein--Hilbert \lagrangian\
of General Relativity, as it should be expected. The scalar--
gravity \lagrangian\ is thus defined (in the absence of matter)
as $\Lphi=L-\Lg$, i.e. $$ \Lphi={1\over 16\pi}\left[(\varphi-
8\pi) R-
{\omega(\varphi)\over\varphi}g^{\mu\nu}\varphi_{,\mu}
\varphi_{,\nu}\right]\sqg\ .\en{2.22} $$ In the presence of
minimally coupled matter the full \lagrangian\ in (2.17)  is thus
decomposed as $L=\Lg+\Lphi+\lmat\sqg$. The term $\Lphi$
generates the variational energy--momentum tensor $$
\tau_{\mu\nu} = {1\over 8\pi}[(8\pi-\varphi)G_{\mu\nu}+
\nabla_{\mu}\nabla_{\nu}\varphi-
g_{\mu\nu}\dal\varphi+{\omega(\varphi)\over\varphi}
(\varphi_{,\mu}\varphi_{,\nu} -\half
g_{\mu\nu}\varphi^{,\alpha}\varphi_{,\alpha})] \en{2.23} $$
and the equation of motion $$ \dal\varphi+{1\over
2\omega}\varphi^{,\alpha}\varphi_{,\alpha}
\left(\tder{\omega}{\varphi}-
{\omega\over\varphi}\right)+{\varphi R\over 2\omega}=0\ .
\en{2.24} $$ The invariance of the action integrals $S_m$ and
$S_{\varphi}$ under spacetime translations generates eight
conservation laws (Noether's theorem, see Appendix A),
$\nabla^{\nu}\tau_{\mu\nu}=0$ and
$\nabla^{\nu}T_{\mu\nu}=0$ (the latter can be directly
verified using (2.24)). $\tau_{\mu\nu}$ explicitly depends on
curvature. In equations (2.23) and (2.24) $g_{\mu\nu}$ is an
external metric field. When the back--reaction of $\varphi$ on
the metric is accounted for, an ambiguity arises. Varying the
action (2.17) with respect to $g_{\mu\nu}$ one arrives at field
equations for spin-two gravity, $$
G_{\mu\nu}(g)=\tau_{\mu\nu}(\varphi,g)+T_{\mu\nu}(\Psi,
g)\ , \en{2.25} $$ and these allow one to eliminate the Einstein
tensor from the expression for $\tau_{\mu\nu}$ and the
curvature scalar from (2.24). After these eliminations the field
equations take the standard form which is usually applied in
SGT theories, $$\eqalignno{ G_{\mu\nu} &=
{1\over\varphi}(\nabla_{\mu}\nabla_{\nu}\varphi-
g_{\mu\nu}\dal\varphi)
+{\omega\over\varphi^2}(\varphi_{,\mu}\varphi_{,\nu}
-\half
g_{\mu\nu}\varphi^{,\alpha}\varphi_{,\alpha})+{8\pi\over
\varphi}T_{\mu\nu}(\Psi,g)\cr &\equiv
\theta_{\mu\nu}(\varphi,g)+{8\pi\over\varphi}T_{\mu\nu}(
\Psi,g)\ , &(2.26)\cr \dal\varphi &={1\over
2\omega+3}\left(8\pi T_{\alpha}{}^{\alpha}-
\tder{\omega}{\varphi}\varphi^{,\alpha}\varphi_{,\alpha}
\right)\ . &(2.27)} $$ By the basic assumption the scalar gravity
does not couple to ordinary matter and its role is confined to
influencing the metric, whereas (2.27) shows that the field
$\varphi$ is influenced by matter although it does not cause a
back--reaction. Actually $\varphi$ interacts only with the metric
field as is seen from (2.24) and (2.25). Furthermore, equation
(2.26) defines the effective stress--energy tensor
$\theta_{\mu\nu}$ for $\varphi$, acting as one of two sources
for the metric field; this tensor is curvature--independent.
Clearly, also $\tau_{\mu\nu}$ is a source of metric gravity,
according to (2.25). Thus, two stress--energy tensors are
assigned to scalar gravity. Such ambiguity arises whenever spin-
-two gravity is generated by two different sources since Einstein
field equations alone determine only the total stress tensor,
which can be expanded in various ways into contributions from
each source separately. The effective stress tensor is not
conserved and the relation $$
\theta_{\mu\nu}=\tau_{\mu\nu}+\left(1-
{8\pi\over\varphi}\right)T_{\mu\nu}\en{2.28} $$ \ssu{yields}
$$
\theta^{\mu\nu}{}_{;\nu}={8\pi\over\varphi^2}T^{\mu\nu}
\varphi_{,\nu}\ .  \en{2.29} $$ It is worth stressing that
$\theta_{\mu\nu}$ appears when the Brans--Dicke scalar acts
as a source of metric gravity, while the variational definition of
$\tau_{\mu\nu}$ is always valid and the latter tensor should
be viewed as the correct expression for a conserved stress-
energy tensor (any possible ambiguities in the construction of
$\Lphi$ are irrelevant in this respect). Whether or not
$\tau_{\mu\nu}$ provides a good physical notion of energy is
a separate problem which will be addressed in Section 3.

In studying conservation laws for matter in Einstein frame we
restrict ourselves to the case of TJBD theory for simplicity,
i.e.~we set $\omega={\it const.}$ Transforming (2.26) and (2.27)
to the Einstein frame one finds the following field equations for
$\g_{\mu\nu}$ and for the redefined scalar
$\phi={1\over\gamma}\ln\varphi$, with
$\gamma=\left(\omega+{3\over2}\right)^{-\half}$:
$$\eqalignno{\G_{\mu\nu} &= \phi_{,\mu}\phi_{,\nu} -\half
\g_{\mu\nu}\g^{\alpha\beta}\phi_{,\alpha}\phi_{,\beta}
+8\pi\exp(-\gamma\phi)T_{\mu\nu}(\Psi,g)\cr &\equiv
t_{\mu\nu}(\phi,\g)+8\pi\exp(-
\gamma\phi)T_{\mu\nu}(\Psi,g)\ , &(2.30)\cr \ndal\phi
&=4\pi\gamma\exp(-
\gamma\phi)T_{\mu\nu}\g^{\mu\nu}\ . &(2.31)} $$ The
notation $T_{\mu\nu}(\Psi,g)$ recalls that the matter stress--
energy tensor is defined in the Jordan frame. By taking the
divergence of (2.30) and using (2.31) one arrives at four matter
conservation laws, $$ \nnabla^{\nu}T_{\mu\nu}-
\gamma\g^{\alpha\beta}(T_{\mu\alpha}\phi_{,\beta}-\half
T_{\alpha\beta}\phi_{,\mu})=0 \en{2.32} $$ Clearly the
massless particles like photons or neutrinos follow null geodesic
worldlines in both frames, while dust moves along timelike
geodesic paths only in the Jordan frame (Appendix B).

Finally, for the sake of completeness\foot{Actually one can also
consider multiscalar--tensor theories of gravity, see [1]}, we
comment on the conformally invariant scalar field model [27].
This theory fits to the general framework of dynamical systems
generated from a general--reativistic scalar field by means of
conformal mappings, and is formally equivalent to a STG theory
via redefinition of the scalar. However, it differs from STG
theories in the physical interpretation given to the scalar field.
The latter is in fact commonly viewed as a special kind of matter
rather than being a spin--zero component of the gravitational
interaction. That the conformally invariant\foot{The conformal
invariance actually means the form--invariance of the field
equation $\dal\chi-{1\over6}R\chi=0$ under arbitrary
conformal map $g_{\mu\nu}\mapsto \Omega^2g_{\mu\nu}$
associated with $\chi\mapsto\Omega^{-1}\chi$ (i.e.~$\chi$ is
scaled like a particle mass).}\ scalar field is equivalent to the
massless linear field under a suitable conformal map is
surprisingly little known to relativists although this fact was
discovered by Bekenstein [28] twenty years  ago\foot{Only
recently has appeared a work [29] applying the theorem.}. The
form of the full action in Jordan frame, $$  S=\int
d^4x\sqg\left[\half R-\half
g^{\mu\nu}\chi_{,\mu}\chi_{,\nu} -{1\over
12}\chi^2R+\lmat(\Psi,g) \right]\ ,\en{2.33} $$ clearly shows
that this frame is assumed to be physical. It is natural to identify
the \lagrangian s for the metric and the scalar as
$\Lg=\half\sqg R$ and $\Lchi=-\sqg(\half
g^{\mu\nu}\chi_{,\mu}\chi_{,\nu}+{1\over 12}\chi^2R)$,
then the latter generates the variational stress--energy tensor for
the scalar, $\tau_{\mu\nu}(\chi,g)$. Instead of deriving it and
the field equations from the \lagrangian, one views (2.33) as a
version of STG theory. Upon comparing (2.33) with (2.17) and
identifying $\varphi=8\pi-{4\pi\over3}\chi^2$, one finds $$
\omega(\varphi)={3\over2}{\varphi\over 8\pi-
\varphi}\qquad \varphi<8\pi\ . \en{2.34} $$ Then the explicit
form of $\tau_{\mu\nu}(\chi,g)$ and the field equations follow
from (2.23)--(2.25). As previously, the Einstein tensor and the
curvature tensor can be eliminated, giving rise to the ``effective
stress-energy'' tensor $\theta_{\mu\nu}$ and to field equations
analogous to (2.26) and (2.27). In the Jordan frame the Noether
theorem implies
$T^{\mu\nu}{}_{;\nu}=\tau^{\mu\nu}{}_{;\nu}=0$, while
(2.28) and (2.29) show that  the effective stress--energy tensor
$\theta_{\mu\nu}$ is not conserved in the presence of
matter\foot{Madsen [30] makes an incorrect statement on the
subject.}, $$ \theta^{\mu\nu}{}_{;\nu}= -12\chi(6-\chi^2)^{-
2}T^{\mu\nu}\chi_{,\nu}\ ,\qquad \chi^2<6  \en{2.35} $$ The
condition $\chi^2<6$ implies $0<\varphi<8\pi$ and for these
ranges of the field variables the theories (2.17) and (2.33) are
equivalent to each other and to (2.21), since the conformal factor
(2.18) is positive. The field $\chi$ is redefined by  $d\phi
\equiv \left(1-{1\over6}\chi^2\right)^{-1}d\chi$, i.e.
$\phi=\sqrt{2\over3}\ln{\sqrt{6}+\chi\over\sqrt{6}-\chi}$.
The inverse transformation back to the Jordan frame is $$
g_{\mu\nu}=2\cosh^2\left({\phi\over\sqrt{6}}\right)
\g_{\mu\nu}\qquad {\rm and}
\qquad\chi=\sqrt{6}\tanh\left({\phi\over\sqrt{6}}\right)
\en{2.36} $$ [For $\chi^2>6$ ($\varphi<0$) the transformation
from the Jordan to the Einstein frame is given by
$\g_{\mu\nu}=-{\varphi\over 16\pi}g_{\mu\nu}$ and
$\phi=\sqrt{2\over3}\ln{\sqrt{6}-\chi\over\sqrt{6}+\chi}$; see
Appendix E].  Thus the conformally invariant field $\chi$ is
nothing but yet another conformal image of the self--gravitating
massless linear scalar $\phi$. In general, a conformal map
$\g_{\mu\nu}=F(\phi)g_{\mu\nu}$ with arbitrary
nonvanishing $F$ transforms the Einstein--Hilbert \lagrangian\
for the scalar $\phi$ minimally coupled to $\g_{\mu\nu}$ into
$$ L(g,\phi) = \sqg\left[F(\phi)R - \left(F-
{3\over2F}F'^2\right)g^{\mu\nu}\phi_{,\mu}\phi_{,\nu}
\right]\ , \en{2.37} $$ which can be transformed by further
conformal maps and field redefinitions into any version of STG
theories (Alonso \etal\ in ref. [11]).

The interrelations between the various gravity theories are
depicted in Figure 1.

We conclude this section by emphasizing that the gravitational
field equations for NLG in the Jordan conformal frame, (2.2) and
(2.9), differ from each other in the matter part, and similarly
equations (2.3) and (2.8) in the Einstein frame have different
matter source term. Thus it is clear that matter dynamics
depends on whether the physical metric is the one forming
either the Jordan or the Einstein (or in any other) frame.
Transforming from the assumed physical frame to the other
(unphysical) one results in a number of bizarre terms depending
on ordinary matter and/or the spin--0 gravity; nevertheless in
each case one finds consistent conservation laws. In particular
we stress that the divergence of the total energy--momentum
tensor (the sum of all the terms in the gravitational field
equations depending on fields different from the metric one)
does not provide a criterion for establishing which metric is
physically acceptable. Accordingly, in the next section we will
revert to a vacuum NLG theory (no ordinary matter), and study
the distinguished role played by energy in gravitational physics;
this will provide a motivation for regarding the Jordan metric as
unphysical.

\section{3. Field redefinitions, physical variables and positivity
of energy}

Having shown that, from the formal viewpoint, NLG and STG
theories can be formulated in either of the two frames without
giving rise to inconsistencies, and having clarified  the
theoretical implications of the choice of the physical metric,  let
us revert to our original question: can one choose the physical
metric on arbitrary grounds?

Brans' response [18] is negative: JBD theory is {\it not\/}
``nothing but'' General Relativity plus a scalar, and the metric
tensor in the Jordan conformal frame has a direct operational
meaning -- test particles move on geodesic worldlines in this
geometry. This is, however, a free assumtion and we have seen
in Section 2 that there is no {\it formal\/} method which could
determine which frame is physical. The problem is further
obscured by the fact that there is no experimental evidence on
interaction of the scalar with known matter. It might therefore
seem (and implicit suggestions are sometimes heard) that a
self--gravitating scalar field can be arbitrarily coupled to the
spacetime metric. Presumably this is the origin of the view that
physics cannot distinguish between conformal frames [16], the
mere fact that the conformal mapping does not affect the
particle mass ratios being clearly insufficient for proving it.

Let us make the terminology more precise. Different
formulations of a theory in different variables (frames) will be
referred to as various versions of the same theory. This includes
not only mere transformations of variables, but also transitions
to {\it dynamically equivalent\/} frames. A theory (expressed
in any version) is {\it physical\/} if there exists a maximally
symmetric ground state solution which is classically stable.
Classical stability means that the ground state solution is stable
against small oscillations -- there are no growing perturbation
modes with imaginary frequencies. A viable physical theory can
be semiclassically unstable: the ground state solution is
separated by a finite barrier from a more stable
(i.e.~lower--energy) state and can decay into it by semiclassical
barrier penetration [31]. The ground state solution may not exist,
e.g.~in Liouville field theory [32], but in gravitational physics
the existence of the ground state solution (Minkowski or de
Sitter space) hardly needs justification.

In most versions of a physical theory it is difficult to establish
whether the ground state solution is stable or not and to extract
its physical contents. Field variables (i.e.~frames) are {\it
physical\/} if they are operationally measurable and if field
fluctuations around the ground state solution, expressed in
terms of these variables, have positive energy density. Since the
energy density cannot be defined for the metric field, the
definition is directly applied to all other fields; the metric tensor
influences the energy density of any matter and thereby the
physical metric is indirectly determined.

Energy density is sensitive to transformations of variables,
particularly to conformal mappings. In terms of unphysical
variables the energy density is indefinite and although the total
energy is formally conserved, it loses most of its practical use.
The ground state solution has (by definition) total energy equal
to zero and when the theory is formulated in the physical
variables (the ``physical version'') the solution represents the
minimum of energy. Thus, stability is closely related to
positivity of energy and instead of searching for growing
perturbation modes one can study the total energy for nearby
solutions [33].

These definitions apply to a relativistic classical field theory
(and not to Newton gravity) and are satisfied by all known
unquantized matter.  {\sl ``For reasons of stability we expect all
reasonable (though not quantum!) field theories to have positive
energy density, and we expect all (classical and quantum) field
theories to have positive mass--energy''\/}  [34]. The Weak
Energy Condition is violated in some quantum states [35] while
for all unquantized matter the Dominant Energy Condition
holds [36]. Whenever the condition is violated one obtains
physically meaningless results [37].

To show how this postulats work, consider a ``ghost'' complex
scalar field in Minkowski space, minimally coupled to
electromagnetism: $$ L=-{1\over
16\pi}F^{\mu\nu}F_{\mu\nu}+D_{\mu}\psi(D^{\mu}\psi)^*
\ ,\qquad
D_{\mu}\psi\equiv\partial_{\mu}\psi+ieA_{\mu}\psi
\en{3.1} $$ Mathematically the theory is acceptable in the sense
that the Cauchy problem for the field equations $$
\partial^{\nu}F_{\mu\nu}-8\pi{\rm Re}(ie\psi^*
D^{\mu}\psi)=0\ ,\qquad D_{\mu}D^{\mu}\psi=0\en{3.2} $$
is well posed. Physically, however, the theory is untenable since
the full energy--momentum tensor $$ T_{\mu\nu}={1\over
4\pi}\left(F_{\mu\alpha}F_{\nu}{}^{\alpha}
-{1\over4}g_{\mu\nu}F^{\alpha\beta}F_{\alpha\beta}\right)
-
\half\left[D_{\mu}\psi(D_{\nu}\psi^*)+D_{\nu}\psi(D_{\mu}
\psi^*) -g_{\mu\nu}D_{\alpha}\psi(D^{\alpha}\psi)^*
\right]\en{3.3} $$ is indefinite. The candidate ground state
solution, $\psi=F_{\mu\nu}=0$, has total energy $E=0$ and is
unstable since any other solution with $F_{\mu\nu}=0$ and
${\rm Im}(\psi)=0$ has negative energy. The ``ground state''
decays via a self--excitation process where energy is pumped
out from the scalar to the electromagnetic field and radiated
away to infinity. Such a system, which can emit an infinite
amount of radiation (while total energy is conserved) is clearly
unphysical.

We shall show that such effects do not occur for NLG theories if
there is exact equivalence between the Jordan and Einstein
frames. On the other hand, if the equivalence breaks down for
some \lagrangian s or solutions, the ground state solution
(Minkowski space) is likely to be unstable. Before doing so, a
few remarks on the problem of the physical (in)equivalence of
the frames and the notion of energy are in order.

A simple example showing that mathematical equivalence
needs not imply a physical one is provided by classical
Hamiltonian mechanics, where the canonical transformation
$P_i=q^i$ and $Q^i=-p_i$, being a mere renaming of positions
and momenta, clearly shows that from the matematical
viewpoint particle positions and momenta do not differ
substantially. On the other hand, to construct operationally
$H(q,p)$ for a given mechanical system one should clearly {\it
discriminate\/} positions and momenta, and kinetic and
potential energy. In this sense, {\it physical\/} positions and
momenta are those that are used to determine, on empirical
and/or theoretical grounds, a Hamiltonian for the given system.
Once $H$ has been constructed in terms of physical variables,
one has the freedom of making arbitrary canonical
transformations. Whether or not the physical variables are
uniquely determined is a separate problem and depends upon
the system under question.

To determine the physical variables and the Hamiltonian (or
Lagrangian) for the system, it  is in general insufficient to study
the system alone: one should take into account its actual and
possible interactions with its surroundings. The greater variety
of interactions, the greater confidence that the dynamical
variables describing the system and its Hamiltonian are
correctly defined. One usually pays less attention to this aspect
since in theoretical physics the physical variables are already
given from empirical data and form a starting point for
theoretical consideration: one is then interested in finding out
the largest group of transformations for a given system, rather
than in restricting the class of allowable frames.

The very possibility that the system can interact with external
agents means that the theory describing it is ``open'', in the
sense that the surrounding in not included in the \lagrangian.
On the contrary, in a ``closed'' theory the system constitutes the
whole universe and no external agent can make an experiment
on it; in this situation, any set of variables describing the system
is equally physical and mathematical equivalence of frames
means the physical one [19]. In a closed theory, in fact, energy is
merely a first integral of motion without the distinguished
features it has in an open theory. Each accepted physical theory
is open in this sense (there are some implicit trends in quantum
gravity to view it as a closed theory [38]) and for instance in
classical electrodynamics and quantum mechanics one has no
doubts which variables have direct physical meaning and which
are merely a convenient mathematical tool for solving a
particular problem.

In the case of gravity, in order to identify the physical variables
and to formulate a physical version of the theory, one should
experimentally study gravitational interactions of various forms
of matter: motion of light and of charges and neutral test
particles, behaviour of clocks and rigid rods etc. The point is that
the presently available empirical data are {\it too scarce\/} to
this aim.

A theoretical criterion for pure gravity or for a system consisting
of gravity and a scalar field is provided by energy, owing to its
unique status in theories of gravity. In no other theories is
energy effectively a charge. In any theory of gravity (including
string--generated ones) the ADM energy should provide a good
notion of energy [39] and the Positive Energy Theorem (see
e.g.~[40]) should hold.

For a NLG theory one should compare energetics in the Jordan
and Einstein frames. There is no generally accepted definition of
gravitational energy for a higher--derivative theory, and in
search for the physical metric one should not compare (as is
usually done) the fourth--order version of the theory (1.2) with
the second--order one. It turns out that the Einstein frame does
{\it not\/} provide the {\it unique\/} second--order version of
the theory, and it was shown in [41] that any NLG theory (as
well as theories with \lagrangian s depending on  Ricci and
Weyl curvatures) can be recast in a version revealing a {\it
formal\/} equivalence with General Relativity without
changing the metric. (Yet the conformal rescaling of the metric is
necessary to have a version of the theory with Einstein--Hilbert
\lagrangian\ (1.6)). This is accomplished with the aid of a \HL\
[41, 42] by applying a Legendre transformation. The \HL\
contains the original unrescaled metric (the Jordan frame), and
to get second--order equations of motion it is necessary to
introduce a new independent field variable, a scalar field $p$.
Thus, the field variables are now $(g_{\mu\nu},p)$, and to
distinguish the frame from the original Jordan frame (consisting
of $g_{\mu\nu}$ alone) it is referred to as Helmholtz Jordan
conformal frame (HJCF).

It should be stressed that the procedure is not an {\it ad hoc\/}
trick and the scalar is (up to possible redefinitions) a canonical
momentum conjugated to the metric $g_{\mu\nu}$, and
represents  an additional degree of freedom existing already in a
NLG theory, in comparison to (vacuum) General Relativity. The
HJCF exists iff $f''(R)\ne0$ (the $R$--regularity condition), i.e.~if
the theory is a truly nonlinear one.

The formalism [41] is outlined in Appendix C. The \HL, which
is dynamically equivalent to (1.1), is $$ \LH=p[R(g)-
r(p)]\sqg+f[r(p)]\sqg\ .\en{3.4} $$ where, as previously, $r(p)$
is a solution of the equation $f'[r(p)]=p$. The resulting field
equations $(C.3)$, after some manipulations, take on the form $$
\eqalignno{ G_{\mu\nu} &= p^{-
1}\nabla_{\mu}\nabla_{\nu}p  -{1\over6}\left\lbrace p^{-
1}f[r(p)]+r(p)\right\rbrace
g_{\mu\nu}\equiv\theta_{\mu\nu}(g,p)&(3.5)\cr
\noalign{\noindent and} \dal p &={2\over3}f[r(p)]- {1\over3}
p\cdot r(p)\ ;&(3.6)}  $$ these are equivalent to (2.14--15) in the
absence of matter. Equations (3.5) have the form of Einstein field
equations with the effective energy--momentum tensor
$\theta_{\mu\nu}$ of the scalar as a source\foot{The fact that
an effective matter source arises from the nonlinear part of the
\lagrangian\ was already noticed and applied in [48] in the case
of a generic quadratic \lagrangian.}. Although the scalar is a
new independent variable, the number of degrees of freedom
remains unchanged: the \lagrangian\ (1.1) describes a system
with 3 degrees of freedom [22, 23] while it is obvious from the
field equations (3.5) and (3.6) that $\LH$ represents two degrees
of freedom for $g_{\mu\nu}$ and one for $p$.

Assuming that the \lagrangian\ $f(R)\sqg$ does not contain the
cosmological term, $f(0)=0$, one finds that Minkowski space is a
possible candidate ground state solution to (1.2)\foot{ We do
not consider in this paper other possible ground states (for
$\Lambda\ne0$), such as de Sitter and anti--de Sitter spaces,
although our arguments can be suitably extended to deal with
them.}. Then $g_{\mu\nu}=\eta_{\mu\nu}$ and $p=p_o$ with
$r(p_o)=R(\eta)=0$ are the solution in the HJCF. Is this solution
stable? To assess it one considers solutions $(g_{\mu\nu},p)$ to
(3.5--6) which approach $(\eta_{\mu\nu},p_o)$ at spatial
infinity at sufficient rate. Then the total energy of these solutions
is given by the ADM surface integral at spatial infinity and the
candidate ground state solution is stable (classically and
semiclassically) if $E_{\rm ADM}\ge0$ and vanishes only for
$(\eta_{\mu\nu},p_o)$. In General Relativity the Positive
Energy Theorem holds provided the r.h.s.~of Einstein equations
satisfies the Dominant Energy Condition for {\it any\/} source
of gravity. The effective stress-energy tensor
$\theta_{\mu\nu}$ for the spin--zero gravity does not satisfy
the condition. The kinetic part of $\theta_{\mu\nu}$ is quite
bizarre: it contains second--order derivatives and is a
homogeneous function of order zero in the field. The latter
peculiarity can be removed by a field redefinition while the
former cannot. In fact, setting $p=F(\varphi)$, with $F'>0$, one
has $\nabla_{\mu}\nabla_{\nu}p=
F'\nabla_{\mu}\nabla_{\nu}\varphi+F''\partial_{\mu}\varphi
\partial_{\nu}\varphi$ and the term survives for any $F$. By
setting $p=\exp(\varphi)$ the effective stress tensor becomes $$
\theta_{\mu\nu}(g,\varphi)=\nabla_{\mu}\nabla_{\nu}
\varphi+\partial_{\mu}\varphi\partial_{\nu}\varphi
-{1\over6}\left\lbrace \exp(-
\varphi)f[r(\exp(\varphi))]+r(\exp(\varphi))\right\rbrace
g_{\mu\nu}\en{3.7} $$ The presence of a linear term in the
effective energy--momentum tensor for any long--range field is
undesired because it causes difficulties in determining the total
energy. The general procedure to couple a field to gravity
consists in regarding it at first as a test field on a fixed spacetime
background (which we choose, for simplicity, to be Ricci--flat),
whereby the energy--momentum tensor can be gauged by any
identically conserved tensor $\sigma_{\mu\nu}$. In the case of
the scalar $\varphi$, the additional term
$\sigma_{\mu\nu}=\nabla_{\mu}\nabla_{\nu}\varphi-
g_{\mu\nu}\dal\varphi$ influences the total energy as
measured at infinity. To avoid such ambiguity it is generally
accepted that the energy--momentum tensor for any field should
be quadratic in the field derivatives and contain no linear terms.
In the case of equation (3.5) this argument is not crucial, because
$\theta_{\mu\nu}$ is uniquely determined by the original
nonlinear \lagrangian\ and no ambiguity can arise. A definitive
argument to regard the linear term in $\theta_{\mu\nu}$ as
unphysical is provided by the fact that due to its presence the
energy density can be of either sign and be transported faster
than light. The appearance of such linear terms signals that
either the theory is unphysical at all (as is the case of Einstein--
Gauss--Bonnet theory with compactified extra dimensions [43])
or merely that the field variables are unphysical and one should
transfer to another frame. The same term also flaws the energy--
momentum tensors (both variational and effective ones) for the
Brans--Dicke scalar and the conformally--invariant scalar field in
the Jordan frame.

In terms of the HJCF field variables one cannot prove that
solutions near the flat space have positive energy; on the
contrary, the indefiniteness of $\theta_{\mu\nu}$ deceptively
suggests that negative--energy solutions exist and Minkowski
space is classically unstable. To establish whether it is actually
stable or not one makes the transformation to the Einstein frame.
While the HJCF exists under the $R$-regularity condition
$f''(R)\ne0$, the Einstein frame exists if the further condition
$f'(R)>0$ is satisfied at all spacetime points for a given solution
(see Appendix E). Since one is interested in solutions which are
asymptotically flat at infinity and do not differ too much (in the
sense given below) from the flat solution in the interior, one
requires $f'(R)>0$ and $f''(R)\ne0$ for $R\rightarrow0$. These
conditions hold for $$ f(R) = R + a R^2 + \sum_{k=3}c_kR^k\
,\qquad{\rm with}\quad a\ne 0\ .\en{3.8} $$ The existence of
the first two terms in the expansion, $R+aR^2$, is of crucial
importance for the equivalence of the Jordan and Einstein
frames. These ensure the invertibility of the Legendre
transformation and yield $p\approx1$ in the vicinity of flat
space. The vicinity consists of all spacetimes for which $R$ is
close to zero; these include spaces of arbitrarily large Riemann
curvature if e.g.~$R_{\mu\nu}=0$. The quadratic \lagrangian\
therefore carries most features of any NLG theory for which the
equivalence holds. On the other hand, if (3.8) does not hold, the
background solution -- with respect to which the ADM energy is
defined -- has no counterpart in the Einstein frame, and
therefore there is no hope that information on the total energy
can be obtained by ordinary general--relativistic techniques.

Let us assume that (3.8) holds. Consider now a spacelike 3--
surface $\Sigma$ embedded in an asympotically flat spacetime
$(M,g_{\mu\nu})$. The asymptotic flatness in the HJCF is
defined as in General Relativity since the field equations are of
second order; the weakest assumptions [44] are (in obvious
notation) $$ g_{\mu\nu}=\eta_{\mu\nu}+O(r^{-\half-
\epsilon})\ ,\qquad g_{\mu\nu,\alpha}=O(r^{-{3\over2}-
\epsilon})\qquad{\rm with}\quad\epsilon>0\ .\en{3.9} $$
Then the leading--order contribution to $R$ is $R=O(r^{-
{5\over2}-\epsilon})$. The HJCF dynamics implies $p=f'(R)$
and the \lagrangian\ (3.8) yields the asymptotic behaviour of
the scalar field, $$ p=1+O(R)=1+O(r^{-{5\over2}-\epsilon})\
.\en{3.10} $$ Under these conditions one now proves, contrary
to what might be expected from the properties of
$\theta_{\mu\nu}$, that the total energy is positive for a NLG
theory. \medskip {\bf Positive Energy Theorem for NLG
Theories.} {\sl Let $\Sigma$ be an asymptotically flat,
nonsigular spacelike hypersurface in a spacetime
$(M,g_{\mu\nu})$ topologically equivalent to $I\!\!R^4$. If:
\item{i)} the \lagrangian\ is given by (3.8);

\item{ii)} $p>0$ and $f''(R)\ne0$ everywhere on $\Sigma$;

\item{iii)} a solution $(g_{\mu\nu},p)$ to (3.5--6) satisfies the
condition (3.9) (and hence (3.10)) on $\Sigma$;

\item{iv)} the potential $V$ (1.7) for the scalar $\phi$ in the
Einstein frame is nonnegative, $V(\phi)\ge0$, on $\Sigma$;

\noindent then the ADM energy in the Jordan frame is
nonnegative, $$ E_{\rm ADM}[g] = \half\int_{S^2_{\infty}}
dS_i(g_{ij,j}-g_{jj,i})\ge0\ .\en{3.11} $$ }

{\bf Proof.} The proof is a direct extension of that given by
Strominger [20] for $f=R+aR^2$. Consider the total energy of the
conformally related solution $(\g_{\mu\nu},\phi)$ in the
Einstein frame. Because of the falloff rate in the HJCF, the metric
$\g_{\mu\nu}=p g_{\mu\nu}$ is asymptotically flat and the
total energy of the solution is finite and given by the ADM
integral over a boundary 2--surface $S^2_{\infty}$ at infinity,
$E_{\rm ADM}[\g] = \half\int_{S^2_{\infty}} dS_i(\g_{ij,j}-
\g_{jj,i})$. If $V\ge0$ the Dominant Energy Condition holds
and this energy is nonnegative [40]. Now, replacing
$\g_{\mu\nu}$ by $g_{\mu\nu}$ and applying (3.10) one
easily finds that $E_{\rm ADM}[\g]=E_{\rm ADM}[g]$ and
hence that $E_{\rm ADM}[g]\ge0$.\hfill{$\dal$}

There is a subtle conceptual difference between our proof and
Strominger's one: the use of the ADM integral for the total
energy is not based on the fact that at large distances the
dynamics is governed by the lower--derivative terms in (1.2),
but rather on the existence of the HJCF. In the latter the integral
is defined as in General Relativity.

 It should be stressed that the global equivalence between the
Jordan and Einstein conformal frames holds if $p>0$ and
$f''(R)\ne0$ everywhere on $(M,g_{\mu\nu})$, while to prove
the theorem one needs only to assume that these conditions hold
on the initial--data surface $\Sigma$, what implies that the
frames are equivalent in some neighbourhood of $\Sigma$.
Consider for example the collapse of a cloud of dust. The exact
relation between $R$ and the dust energy density $\rho$
depends on the form of the field equations (which metric is
physical, Section 2); in general one expects that $R$ grows when
$\rho$ does. On the initial surface $\Sigma$ the dust is diluted
and $R\approx0$, thereby the frames are equivalent in the
vicinity of $\Sigma$. Near the singularity $\rho$ is divergent
and $R$ is unbounded, $p$ may change sign and thus in this
region of the manifold there may be no mapping to the Einstein
frame. Yet the energy, being conserved, is still given by its value
on $\Sigma$.

In the class of solutions for which the two conformal frames are
at least locally equivalent, Minkowski space is the unique one
having zero energy. In fact, let $M$ contain a spacelike surface
$\Sigma$ on which the assumptions of the theorem hold. Then,
they hold in some neighbourhood ${\cal U}(\Sigma)$ and the
spacetime region $({\cal U}(\Sigma),g_{\mu\nu})$ can be
mapped onto $({\cal U}(\Sigma),\g_{\mu\nu})$. Let $E_{\rm
ADM}[g]=0$ when evaluated on $\Sigma$; then
$\g_{\mu\nu}$ is flat in ${\cal U}(\Sigma)$ (Minkowski
spacetime is the unique zero-energy solution in General
Relativity) and the solution
$(\g_{\mu\nu}=\eta_{\mu\nu},\phi=0)$ is identically
mapped back onto $(g_{\mu\nu}=\eta_{\mu\nu},p=1)$ in
${\cal U}(\Sigma)$. Then the spacetime $(M,g_{\mu\nu})$ is
flat in the open region ${\cal U}(\Sigma)$ and this solution can
be analytically extended onto the entire manifold $M$, thereby
the spacetime actually is Minkowski space ($I\!\!R^4$ topology
is always assumed).

Thus the total energy is positive in the Jordan frame for all
solutions to (1.2) containing an open region $\cal U(\Sigma)$
where these are close to Minkowski space ($R$ close to zero)
and where $V(\phi)\ge0$. The flat solution is classically stable
against decay into these solutions and {\it is\/} a ground state
solution for a given NLG theory.

Now we consider a few examples.

\item{\bf (1)} $f(R)=R+aR^2$. The \lagrangian\ is $R$--
regular, since $f''=2a\ne0$ everywhere.

\itemitem{(A)} $a>0$. This is the case studied by Strominger
[20]. Using the variable $p$ instead of $\phi$ one has
$V(p)={1\over 8ap^2}(p-1)^2\ge0$ everywhere, while $p>0$ for
$R>-{1\over 2a}$. For all solutions in the Jordan frame such that
there exists a spacelike asymptotically flat surface $\Sigma$
with $R>-{1\over 2a}$ on it, the total energy is nonnegative and
flat space is classically stable. Strominger has also shown that
Minkowski space is the unique solution in the Jordan frame
with vanishing energy.

\itemitem{(B)} $a<0$. The conformal equivalence holds in the
region $R<|{1\over 2a}|$, but $V(p)\le0$ everywhere and the
theorem does not hold. One may expect that in the vicinity of
flat space there are solutions with negative energy and
Minkowski space is classically unstable.

\item{\bf (2)} $f(R)={1\over a}(\exp(aR)-1)$. This is again a
$R$--regular \lagrangian. Furthermore, the Jordan and Einstein
frames are globally equivalent since $p=\exp(aR)>0$ for any
solution. The potential $V(p)={1\over2ap}(\ln p+{1\over p}-1)$
is positive for $a>0$ and negative for $a<0$ ($V(1)=0$). Thus for
$a>0$ {\it all\/} asymptotically flat solutions have positive total
energy and flat space is stable against {\it any\/} perturbations,
small or large (classical and semiclassical stability). On the
contrary, for $a<0$ the Dominant Energy Condition does not
hold, what signals existence of negative--energy states and
classical instability of flat space.

\item{\bf (3)} $f(R)={1\over a}\ln(1+aR)$. In this case one has
$f''(R)=-{a\over(1+aR)^2}\ne0$ for $R\ne -{1\over a}$, and
$V(p)={1\over2ap^2}(\ln p-p+1)$. For $a>0$ one finds
$V(p)<0$ for $p\ne1$ and flat space can classically decay into
negative--energy states. For $a<0$ the conformal equivalence
holds for $R<-{1\over a}$ and $V(p)>0$ for all $p\ne1$, thus for
these solutions the total energy is positive and flat space is
classically stable.

\noindent In all these examples states close to Minkowski space
have positive energy if the coefficient of the $R^2$ term in the
Taylor expansion of $f$ is positive. This is a generic feature of
\lagrangian s (3.8). For $R$ close to zero the two frames are
equivalent and the theorem holds iff the potential is positive.
For vacuum the relation $r(p)=R(g)$ holds; using it in the
expression for $V(p)$, given after equation (2.15), and
expanding the potential around $R=0$ one arrives at $V=\half a
R^2+O(R^3)$ for solutions. All $R=0$ solutions represent a local
extremum of the potential. Hence energetics and stability of flat
space are determined by the $R^2$ term in the \lagrangian\ of
a NLG theory.

The lowest--order contribution $R+aR^2$ to the full
\lagrangian\ is crucial for {\it classical\/} stability of
Minkowski space. Whenever these terms are absent, e.g.~for
$f=R^k$, $k>2$, the conformal equivalence with the Einstein
frame version is broken at flat space. In this case as well as when
the Einstein frame is defined for $R=0$ but the potential is
negative on $\Sigma$, the classical decay of flat space may
occur because the Positive Energy Theorem at Null Infinity (see
[40]) does not hold. While the ADM energy is conserved, there
may be metric perturbations about the background such that
gravitational waves carry away unbound amounts of energy
and the Bondi--Sachs mass remaining in the system decreases to
minus infinity.

Whenever the ADM mass is positive for solutions close to flat
space (i.e.~if $a>0$), the latter need not be {\it
semiclassically\/} stable. There may exist solutions (in the
Jordan frame) with negative energy which are far from
Minkowski space. These are spacetimes so highly curved (large
$|R|$) in the interior, that either the equivalence with the
Einstein frame is violated ($p$ changes sign on each
asymptotically flat $\Sigma$) or the potential becomes negative
in a region of any $\Sigma$. The occurrence of any of the two
possibilities depends on the form of the \lagrangian, e.g.~for the
exponential one (example  2 above) there are no such solutions,
while for a cubic \lagrangian, $R+aR^2-b^2R^3$, the potential
is negative for large $p$ and some values of $a$. Semiclassical
instability of standard ground state solutions is a generic feature
of higher--dimensional theories: it was found both in
Kaluza--Klein theory in five [31, 45] and in ten [46] dimensions,
as well as in string theory [46]. Although disturbing in itself, it is
not dangerous. Classical instability is a more severe problem
and usually makes the theory untenable. If there is another
possible ground state which is stable, however, it is harmless, as
in the case of the uncompactified Einstein--Gauss--Bonnet
theory, where $d$--dimensional anti--de Sitter space is unstable
while Minkowski space is stable [47]. Minkowski space is
classically unstable in a gravity theory with \lagrangian\
$aR+bR^2+cC_{\alpha\beta\mu\nu}C^{\alpha\beta\mu\nu}
$ [48] and in a four--dimensional theory resulting from Einstein-
-Gauss--Bonnet theory with compactified extra dimensions [43]
(in all these cases there is no {\it conformal\/} mapping of the
theory onto a model in General Relativity).

Equality of the total energy in the two conformal frames is due
to the fact that $E_{\rm ADM}$, being effectively a charge, is
evaluated at spatial infinity (where $p\rightarrow1$) and is
rather loosely related to the interior. The only detailed
information about the interior of a gravitating system that is
needed is whether all local matter energy--flow vectors are
timelike or null. This connection is lost in the Jordan frame. Also
the Bondi--Sachs mass, when expressed in terms of the Jordan
frame, does not satisfy the correct evolution equations, as is
mentioned in [1]. These flaws do not reflect the genuine
properties of spin--0 and spin--2 gravity, these are merely due to
an improper choice of field variables. The effective stress tensor
$\theta_{\mu\nu}$ (3.7) does {\it not\/} provide the true
physical energy--momentum of the scalar field. As any
redefinition of the scalar cannot help, it is the metric variable
which needs redefinition.

Accordingly, the Jordan frame should be regarded as unphysical
not by an arbitrary choice of definition but because in the
second--order version of the theory -- whereby the spin--0 and
spin--2 component of gravity are singled out -- these variables
are unrelated to the total energy of the system. On the other
hand, the Einstein frame satisfies all general requirements of
relativistic field theories. Therefore it is the physical frame, and
$\g_{\mu\nu}$ determines the spacetime intervals in the real
world. It provides the physical version of a NLG theory. All
other conformally--related versions of the theory, including the
initial one, are unphysical in some aspects\foot{The need of
making all calculations in the Einstein frame for STG theories is
stressed in [1],  [8] and [9], although these authors assume the
Jordan frame as physical.}. This physical frame is {\it
uniquely\/} determined by the \lagrangian\ (1.1) and the
physical metric for the vacuum theory is
$\g_{\mu\nu}=f'(R)g_{\mu\nu}$. Physically, this
transformation means a transition to the frame where all fields,
including spin--0 gravity, satisfy the Dominant Energy
Condition. Then it is possible to establish that Minkowski space
is a classically stable local minimum of the energy. This holds
true, however, if the potential of the scalar field does not attain a
local maximum at flat space, which would destabilize the latter.
The only remedy is to exclude as unphysical all \lagrangian s
giving rise to negative potentials near flat space, and this
amounts to the requirement $a>0$ in (3.8).

\section{4. Conclusions}

Our analysis of the physical features of energy in NLG theories
leads to the conclusion that the Einstein--frame metric should be
regarded as the physical one. In this perspective, whenever a
nonlinear \lagrangian\ $L$ admits flat space as a stable ground
state solution, it is {\it physically\/} equivalent to a scalar field
with a potential $V$ (determined by $L$) minimally coupled to
the rescaled metric $\g_{\mu\nu}$ in General Relativity. Then,
any form of matter should minimally couple to $\g_{\mu\nu}$
and not to the scalar. The Strong Equivalence Principle is then
retained and the scalar, which can (though not necessarily) be
viewed as a spin--0 component of universal gravity, appears
only as a contribution to the full source in the Einstein field
equations.

Scalar fields are now very popular in cosmology. Instead of
introducing {\it ad hoc\/} a special form of the scalar field
potential in order to solve a particular problem, one can start
from some nonlinear \lagrangian\ which provides the desired
potential, provided that one can give a deeper motivation for the
former. Our conclusion does not mean that any viable nonlinear
gravity theory is identical with General Relativity. Although the
scalar field has no impact on motion of ordinary matter (in
particular, free test particles move on geodesic lines), the space
of solutions is different. For instance, for nonconstant scalar field
there are no black holes with regular event horizons since a
black hole cannot have a scalar hair [49].

It is amusing to notice that twenty years ago Bicknell [2]
concluded his study of the phenomenology of purely quadratic
theories with the words: {\sl ``These results eliminate
gravitational theories based on quadratic Lagrangians from the
realm of viable gravitational theories and point strongly
towards the uniqueness of the Einstein equations''}. Our results
regarding the \lagrangian s giving rise to potentials which
attain a maximum for flat space ($a<0$) suggest the some
conclusion. However, a rigorous proof of the instability of flat
space in that case is not yet available. The case of \lagrangian s
for which a conformal mapping onto the Einstein frame does not
exists at Minkowski space (e.g.~$R^3$) is more subtle. The
presently known techniques of investigating energy and
stability fail there and it remains an open problem whether
positivity of energy can be shown for a subclass of these
\lagrangian s. Another open problem is how to deal with
\lagrangian s which are not of class $C^3$ at $R=0$,
e.g.~$R+aR^2+R^{7\over3}$.

Finally, it should be emphasized that our method of
determining the physical metric for a gravity theory, by
studying the energy and stability in the vacuum theory, does
not apply when the interaction of matter with gravity is already
prescribed by a more fundamental theory. This is the case,
e.g.~of string--generated gravity, where the string action in the
four--dimensional field--theory limit contains a number of
fundamental bosonic fields coupled to the metric fiels and to the
dilaton. The problem of identifying the physical metric appears
there too [50] and should be dealt with in a separate way.

\section{Acknowledgements}

We are deeply grateful to Marco Ferraris, Mauro Francaviglia,
Andrzej Staruszkiewicz and Igor Volovich for valuable and
helpful comments and discussions. L.M.S. acknowledges the
hospitality of the ``J.--L.~Lagrange'' Institute of Mathematical
Physics at Turin, where this work was done. This work was
supported by G.N.F.M. (National Group for Mathematical
Physics) of Italian C.N.R. and by a grant of the Polish
Committee for Scientific Research.

\vfill\eject

\section{Appendix A: Generalized Bianchi Identity and
Conservation Laws}

Although it is well known to the expert in gravitational physics,
we provide here, for the reader's convenience, the derivation of
the Noether conservation laws for gravitating matter, based on a
generalized Bianchi identity. Consider a generic action for
gravity and a matter field $\Psi$: $$\eqalignno{ S = S_g + S_m
=& \int_{\Omega}d^4x\sqg
F(g_{\alpha\beta},R_{\alpha\beta\rho\sigma},
\nabla_{\mu}R_{\alpha\beta\rho\sigma},\ldots,
\nabla_{\mu_1}\ldots\nabla_{\mu_n}
R_{\alpha\beta\rho\sigma})\cr
&+\int_{\Omega}d^4x\sqg\lmat(g;\Psi)\ ;&(A.1)} $$ here $F$
is a scalar function depending on the metric and its derivatives
(via the curvature tensor) up to $(n+2)$th order. One takes an
infinitesimal point transformation,  $x^{\mu}\mapsto
x'^{\mu}=x^{\mu}+\epsilon\xi^{\mu}(x)$, with $\xi^{\mu}$
being an arbitrary vector field which vanishes, together with all
its derivatives up to $(n+2)$th order, at the boundary of the
region $\Omega$. The action integrals for gravity and matter,
$S_g$ and $S_m$, are separately conserved under the
transformation. The variation of the gravitational action is $$
\delta S_g=0=\int_{\Omega}d^4x\delta(\sqg F)\ . \en{A.2} $$
By applying $(n+2)$ times the Gauss theorem and discarding
the surface integrals one arrives at the standard formula $$
\delta S_g=\int_{\Omega}d^4x\vder{(\sqg
F)}{g^{\mu\nu}}\delta g^{\mu\nu}\ , \en{A.3} $$ \su{where}
$$\eqalignno{ \vder{(\sqg F)}{g^{\mu\nu}}\equiv &
\pder{(\sqg F)}{g^{\mu\nu}}  -
\partial_{\alpha}\left(\pder{(\sqg F)}{g^{\mu\nu}_{\
,\alpha}}\right) +
\partial_{\alpha}\partial_{\beta}\left(\pder{(\sqg
F)}{g^{\mu\nu}_{\ ,\alpha\beta}}\right)\cr &+\cdots+ (-
1)^{n+2}
\partial_{\alpha_1}\cdots\partial_{\alpha_{n+2}}\left(\pder{(
\sqg F)}{g^{\mu\nu}_{\
,\alpha_1\cdots\alpha_{n+2}}}\right)\ . &(A.4)} $$ One
replaces the tensor densities by pure tensors according to $$
\vder{(\sqg F)}{g^{\mu\nu}}\equiv \sqg Q_{\mu\nu}\
,\qquad Q_{\mu\nu}=Q_{\nu\mu}\ ,\en {A.5} $$ \su{then} $$
\delta S_g=\int_{\Omega}d^4x \sqg Q_{\mu\nu}\delta
g^{\mu\nu}\ . \en{A.6} $$ For the case of the infinitesimal
point transformation one finds that $\delta
g^{\mu\nu}=\epsilon(\xi^{\mu;\nu}+\xi^{\nu;\mu})$, what
ensures that the surface integrals in $\delta S_g$ vanish.
Furthermore, $$ \eqalign{0&=\delta S_g
=\epsilon\int_{\Omega}d^4x\sqg
Q_{\mu\nu}(\xi^{\mu;\nu}+\xi^{\nu;\mu})=
2\epsilon\int_{\Omega}d^4x\sqg
Q_{\mu\nu}\xi^{\mu;\nu}\cr = &
2\epsilon\int_{\Omega}d^4x\sqg[\nabla^{\nu}(Q_{\mu\nu}
\xi^{\mu})-\xi^{\mu}\nabla^{\nu}Q_{\mu\nu}]=
2\epsilon\int_{\partial\Omega}Q_{\mu\nu}\xi^{\mu}
dS^{\nu} -2\epsilon\int_{\Omega}d^4x\sqg\xi^{\mu}
Q_{\mu\nu}{}^{;\nu}. } $$ The surface integral vanishes, and
due to the arbitrariness of the vector $\xi^{\mu}$ inside the
integration domain $\Omega$ one gets
$Q_{\mu\nu}{}^{;\nu}=0$. This is an identity which holds for
any metric, independently of the field equations. In the case
$F=f(R)$ the tensor $Q_{\mu\nu}$ is equal to the l.h.s.~of
equation (1.2). Any scalar function $F$ of the spacetime metric
and its partial derivatives up to $(n+2)$th order gives rise to a
generalized Bianchi identity $$
\nabla^{\nu}\left({1\over\sqg}\vder{(\sqg
F)}{g^{\mu\nu}}\right)=0\ .\en{A.7} $$ The symmetric tensor
$Q_{\mu\nu}$ is a function of the metric and its derivatives up
to $(2n+4)$th order, unless $F$ is a linear combination (with
constant coefficients) of particular expressions like $R$
(Einstein--Hilbert \lagrangian),
$R_{\alpha\beta\rho\sigma}R^{\alpha\beta\rho\sigma}-
4R_{\alpha\beta}R^{\alpha\beta}+R^2$ (Gauss--Bonnet term),
etc. (which correspond to the Euler--Poincar\'e topological
densities in dimension two, four, and so on). In other terms, we
can construct out of the Riemann tensor, besides the well-known
contracted Bianchi identity  $G_{\mu\nu}{}^{;\nu}=0$, an
infinite number of differential identities (which can all be
derived from the ordinary Bianchi identity and its higher
derivatives), sometimes called {\it strong conservation laws}.

Upong variating the action $(A.1)$ with respect to arbitrary
variations $\delta g^{\mu\nu}$ of the metric one arrives at the
gravitational field equations $$ Q_{\mu\nu}(g)=\half
T_{\mu\nu}(g,\psi)\ . \en{A.8} $$ The matter stress--energy
tensor, defined by $\sqg T_{\mu\nu}=-
2\vder{}{g^{\mu\nu}}(\sqg\lmat)$ may depend on first and
second derivatives of the metric. The coefficient $\half$ in
$(A.8)$ is due to the normalization of the gravitational
\lagrangian\ assumed in $(A.1)$. Actually, in order to have
correspondence to General Relativity, upon expanding $F$ in
powers of the curvature scalar and other curvature invariants, if
the linear term $R$ has a coefficient $a$, then the standard
matter \lagrangian\ $\lmat$ in $(A.1)$ should have coefficient
$2a$ (having set $8\pi G=1$). Assuming that the gravitational
field equations hold, one takes the divergence of $(A.8)$, and
using the identity $(A.7)$ one gets four conservation laws for
matter, $T_{\mu\nu}{}^{;\nu}=0$.

An equivalent and somewhat shorter way of derivation of the
conservation laws consists in utilizing the invariance of the
matter action $S_m$ under infinitesimal point transformations.
These transformations give rise to the variations $\delta
g^{\mu\nu}=2\epsilon\xi^{(\mu;\nu)}$ and $\delta\Psi$.
Assuming that the variation $\delta\Psi$ and its relevant
derivatives vanish at the boundary $\partial\Omega$ of the
integration domain, one finds that $$ \delta
S_m=0=\int_{\Omega}d^4x\left[
\vder{(\sqg\lmat)}{g^{\mu\nu}}\delta g^{\mu\nu}+
\vder{(\sqg\lmat)}{\Psi}\delta\Psi\right]\ . \en{A.9} $$
Provided that the equations of motion for matter hold,
$\vder{(\sqg\lmat)}{\Psi}=0$, the invariance of the action
yields $$ \delta S_m=-\half\int_{\Omega}d^4x\sqg
T_{\mu\nu}\delta g^{\mu\nu} =-
\epsilon\int_{\Omega}d^4x\sqg T_{\mu\nu}\xi^{\mu;\nu}
=\epsilon\int_{\Omega}d^4x\sqg
\xi^{\mu}T_{\mu\nu}{}^{;\nu}\ , \en{A.10} $$ hence
$T_{\mu\nu}{}^{;\nu}=0$. These conservation laws are actually
either equivalent to the equations of motion for matter or to a
subset of them (the number of equations of motion may exceed
four).

\section{Appendix B: Motion of dust and photons in the Jordan
and Einstein conformal frames}

We consider here the explicit forms of the conservation laws in
the cases I and II for NLG theories and in the Einstein frame for
STG theories. We show on explicit examples that the different
versions of conservation laws reflect the known geometrical
property that null geodesic worldlines are preserved under
conformal rescaling, while timelike geodesic paths in one frame
correspond to non--geodesic paths in the other frame: this fact
has a physical counterpart in the different coupling between
matter and scalar field in the two frames. The consistency of the
conservation laws described in Section 2 with the physical
assumptions is thus ensured both in case I and in case II.

 \subsection{Case I: the Jordan frame is physical.}

In Einstein frame the conservation laws for matter are given by
(2.8). Inserting the explicit form of the $\phi$--field stress tensor
$t_{\mu\nu}$ and making use of the equation of motion (2.6)
one arrives at four equations $$
\g^{\alpha\beta}\left[\nnabla_{\alpha}T_{\mu\beta}+{1\over
\sqrt{6}}T_{\alpha\beta}\phi_{,\mu}- \sqrt{2\over3}
T_{\mu\beta}\phi_{,\alpha}\right]=0 \en{B.1} $$ Clarly, as is
seen from the \lagrangian s (2.1) and (2.3), only conformally
invariant matter does not interact with the scalar (for this matter
$\g^{\alpha\beta}T_{\alpha\beta}=0$ and the equation of
motion for $\phi$ becomes $\Psi$--independent). This is the
case of electromagnetic field, hence photons travel along null
geodesic lines both in the physical and in the rescaled
geometries. We show that this result also follows from the
conservation laws $(B.1)$.

Consider the null eletromagnetic field which is the curved--
space classical representation of the photon, $$ F_{\mu\nu} =
k_{\mu}q_{\nu}-k_{\nu}q_{\mu}\ ,\qquad {\rm with}\qquad
k^{\mu}k_{\mu}=k^{\mu}q_{\mu}=0\ ; \en{B.2} $$ here,
$k_{\mu}$ are the covariant components of the wave vector and
$q_{\mu}$ is a spacelike covector determined up to
$q_{\mu}\mapsto q_{\mu} + \lambda k_{\mu}$. Since
$F_{\mu\nu}=\partial_{\mu}A_{\nu}-\partial_{\nu}A_{\mu}$
is metric--independent, the components $k_{\mu}$ and
$q_{\mu}$ are independent of the conformal rescaling of the
metric. Introducing in Jordan conformal frame a scalar
$w={1\over 4\pi}g^{\alpha\beta}q_{\alpha}q_{\beta}>0$ one
finds that the stress tensor for the null field is given in this frame
by  $T_{\mu\nu}(F,g)=w\,k_{\mu}k_{\nu}$. Then
$\nabla_{\nu}T^{\mu\nu}=0$ imply that
$k^{\nu}\nabla_{\nu}k^{\mu}\propto k^{\mu}$ -- the wave
vector is tangent to null geodesic curves.

In transforming to the Einstein frame, $g_{\mu\nu}\mapsto
p\,g_{\mu\nu}$, $k_{\mu}\mapsto k_{\mu}$ and
$q_{\mu}\mapsto q_{\mu}$, thus $T_{\mu\nu}(F,p^{-
1}\g)=p\w\,k_{\mu}k_{\nu}$, with $\w={1\over
4\pi}\g^{\alpha\beta}q_{\alpha}q_{\beta}$. Equations $(B.1)$
for this stress--energy tensor reduce to $$
k^{\nu}\nnabla_{\nu}k^{\mu}=\left[\sqrt{2\over3}\phi_{,\nu
}k^{\nu}-\nnabla_{\nu}k^{\nu} -k^{\nu}\partial_{\nu}
\ln(p\w)\right]k^{\mu} $$ and the wave vector remains
tangent to null geodesics but the parameter $\sigma$ defined
by $k^{\nu}=\tder{x^{\nu}}{\sigma}$ is not the affine
parameter along them.

On the other hand, massive particles do interact with the
$\phi$--field. Consider a self--gravitating dust. In Jordan frame
its stress tensor is $T_{\mu\nu}=\rho u_{\mu}u_{\nu}$ and
$\nabla_{\nu}T^{\mu\nu}=0$ gives rise to motion along
timelike geodesics, $u^{\nu}\nabla_{\nu}u^{\mu}=0$. Under
the conformal rescaling one finds that $u_{\mu}=p^{\half}
\tu_{\mu}$ and $\rho=p^2\nrho$ [28], thus the dust stress
tensor defined in the Jordan frame and expressed in terms of the
Einstein frame is $T_{\mu\nu}=p\nrho \tu_{\mu}\tu_{\nu}$.
While in the Jordan frame the density current is conserved,
$\nabla_{\nu}(\rho u^{\nu})=0$, in Einstein frame the
corresponding density current is coupled to the scalar, $$
\nnabla_{\nu}(\nrho\tu^{\nu})=-
{1\over\sqrt{6}}\nrho\tu^{\nu}\phi_{,\nu}\ .\en{B.3} $$
Using this fact in the conservation laws $(B.1)$ one arrives at the
following equations of motion for dust in $\g_{\mu\nu}$
metric: $$ \tu^{\nu}\nnabla_{\nu}\tu^{\mu}={1\over\sqrt{6}}
(\phi^{,\mu}+\phi_{,\nu}\tu^{\mu}\tu^{\nu})\ .\en{B.4}  $$
In particular, the dust worldlines are geodesic for
$\g_{\mu\nu}$ only if $\phi_{,\mu}=0$.

\subsection{Case II: the Einstein frame is physical.}

Consider a pressureless dust either being a source in the field
equations (2.10) or just a cloud of test particles. In the Einstein
frame its energy--momentum tensor is divergenceless and the
particles follow timelike geodesic curves. In MJCF the
interaction with gravity becomes highly complicated (equation
$(C.19)$ in Appendix C), and the worldlines are not geodesic.

Massless non--interacting particles move along null geodesic
paths in the physical metric $\g_{\mu\nu}$ and this property is
preserved under any conformal rescaling. This fact can be also
explicitly derived from the matter conservation laws valid in
MJCF. For conformally invariant matter\foot{In this case the
MJCF coincides with the VJCF.}\ the field equations $(C.19)$
reduce to $(2.13)$ and the generalized Bianchi identity gives rise
to the following four conservation laws:   $$
\nabla^{\nu}[f'(R)T_{\mu\nu}(\Psi,f'(R)g_{\alpha\beta})]=0\
.\en{B.5} $$ Thus the proof essentially follows the same lines as
in Case I.

\subsection{Scalar--tensor gravity theories}

By assumption, in the Jordan frame dust particles follow
timelike geodesic worldlines and photons follow the null ones.
In the Einstein frame, we can relate the motion of matter to the
conservation laws (2.33) by the same procedure used for NLG
theories, Case I; in fact, equation $(B.1)$ coincides with (2.33)
with $\omega\equiv 0$. The calculations for the general case
follow essentially the same lines, leading to the conclusion that
photon worldlines are geodesic in both frames while dust
particles move along nongeodesic curves for the metric
$\g_{\mu\nu}$: $$
\tu^{\nu}\nnabla_{\nu}\tu^{\mu}=\half\gamma(\phi^{,\mu
}+\phi_{,\nu}\tu^{\mu}\tu^{\nu})\ .\en{B.6}  $$

\section{Appendix C: Direct and Inverse Legendre
Transformations in the Presence of Matter}

The equivalence of any NLG theory (1.1) to General Relativity
plus the scalar field (1.6) is dynamical in the sense that the
spaces of {\it classical solutions\/} for each theory are locally
isomorphic (see Appendix E), the isomorphism being given by
(1.3). This is directly achieved by the conformal transformation
of the field equations (1.2) [2, 5], and is sufficient for various
purposes. On the other hand, using this {\it ad hoc\/}
procedure it is rather difficult to establish the particle contents of
NLG theories [22, 23] and to recognize the field $\phi$ as an
independent degree of freedom present in the theory; instead
one merely views $\phi$ as a function of the curvature scalar
$R$ of the original metric [6]. These goals can be easily achieved
using a \lagrangian\ formalism. A Legendre transformation
allows to transform the NLG \lagrangian\ (1.1) into a
dynamically equivalent one, linear in the curvature scalar and
including an auxiliary scalar field. This equivalent \lagrangian\
is named ``\HL'' by analogy with classical mechanics. Besides
being elegant and mathematically well--grounded, the Legendre
transformation and the related \HL\ turn out to be
indispensable tools in dealing with NLG theories: one would
hardly guess how matter fields with a prescribed coupling with
the Einstein--frame metric interact with the Jordan--frame
metric, without relying on the corresponding \lagrangian s. As
is shown in Section 2, a na\"\i ve approach leads to
inconsistencies.

Being linear in the second derivatives of the metric tensor, the
\HL\ generates second--order gravitational equations; then the
Einstein--frame \lagrangian\ (1.6) is obtained   by re--
expressing the former in terms of the conformally rescaled
metric $\g$. It is natural then to use the \HL\ to compare the
dynamics in the two frames, and therefore it is a central
ingredient in our approach. In this Appendix we outline of the
general setting of the method [3, 4, 41] (we refer the reader to
[51] for a rigorous mathematical exposition), and then present in
detail how matter interaction terms should be inserted into the
gravitational \lagrangian s (both the original nonlinear one and
the \HL) when the matter fields are assumed to be minimally
coupled to the Einstein--frame metric.

\subsection{C.1 \HL.}

For the nonlinear lagrangian density (1.1) one introduces the {\it
generalized conjugate momentum}  $$ p =
{1\over\sqg}{\partial L\over\partial R}\ .\en{C.1} $$ The
scalar field $p$ is clearly nothing else but the scalar defined by
(1.3). As in Section 1, let $r(p)$ be a  function such that
$f'[R]\big|_{R=r(p)}\equiv p$; such a function exists if
$f''(R)\ne 0$, and in this case we say that the nonlinear
\lagrangian\ (1.1) is {\it $R$-regular}. The \HL\ corresponding
to (1.1) is then defined as $$ \LH=p[R(g)-r(p)]\sqg+f[r(p)]\sqg
.\en{C.2} $$ The action functional corresponding to $(C.2)$ is
formally a degenerate case of the STG action (2.7), with
$\omega\equiv 0$ and nontrivial cosmological function. A
basic feature of the \HL\ is that it does not contain derivatives
of the scalar field $p$. The Euler--Lagrange equations obtained
by varying the metric $g_{\mu\nu}$ and the scalar $p$ {\it
independently\/} in the action defined by $(C.2)$ are  $$
\left\lbrace\matrix{ \hfill R(g) &=& r(p)\hfill\cr\cr
pG_{\mu\nu}(g) &=& \nabla_{\mu}\nabla_{\nu}p-
g_{\mu\nu}\dal p +\half\big(f[r(p)]-p\cdot
r(p)\big)g_{\mu\nu}}\right. \en{C.3} $$ which are manifestly
equivalent to (1.2).

What has been done is nothing but a generalization of a
standard method of classical mechanics. Given a first--order
Lagrangian $L=L(t,q^i,\dot q^i)$ for a system of point particles,
the second--order Euler--Lagrange equations ${d\over
dt}{\partial L\over\partial \dot q^i}-{\partial L\over\partial
q^i}=0$ can be recast into a first--order system, provided the
Lagrangian is regular ($\det\left|{\partial L\over\partial
q^i\partial q^j}\right|\ne 0$). This can be done in two ways:
either one writes the Euler--Lagrange equations in normal form,
$\ddot q^i=a^i(t,q^j,\dot q^j)$, then one introduces
independent velocity variables $u^i$ and writes the equivalent
system in the velocity space $$ \left\lbrace\matrix{\dot
q^i=u^i\hfill\cr \dot u^i=a^i(t,q^j,u^j)}\right.\en{C.4} $$ or
one defines the Legendre map
$(q^i,u^i)\mapsto(q^i,p_i={\partial L\over\partial\dot q^i})$
and finds the inverse map $u^i=u^i(q^j,p_j)$, which transforms
the previous first--order system into the Hamilton equations in
the phase space of the system. An interesting point, which is not
always emphasized in textbooks, is that both first--order
systems arise from a variational principle, defined by a
degenerate first--order Lagrangian, the \HL\ [42] $$
\LH=p_i\left(\dot q^i-u^i\right)+L(q^j,u^j)\ , $$ which can be
regarded either as a function over the velocity space, whereby
$(q^j,u^j)$ are the dynamical variables and $p_i=p_i(q^j,u^j)$,
or as a function over the phase space, where the variables are
$(q^j,p_j)$ and $u^i=u^i(q^j,p_j)$ is the inverse of the Legendre
map defined above. In fact, the solutions of $(C.4)$ are the
extremal curves in the velocity space for the action functional $$
S_{_L} = \int \left\lbrace p_i(q^j,u^j)\left[\dot q^i-
u^i\right]+L(q^j,u^j)\right\rbrace dt\ , $$ while the Hamilton
equations can be derived from the action (in phase space) $$
S_{_H} = \int \left\lbrace p_i\left[\dot q^i-
u^i(q^j,p_j)\right]+L[q^j,u^j(q^k,p_k)]\right\rbrace dt. $$  The
\lagrangian\ $(C.2)$ plays the role of the \HL\ in phase space
for any $R$-regular NLG \lagrangian\ (1.1). As in particle
mechanics, it is also possible to define a \HL\ in the velocity
space: this corresponds to regarding directly the ``velocity''$R$
as an independent field $u$ (as is done e.g.~in [6], following [52]
and without invoking the notion of \HL), and introducing the
\lagrangian:  $$ \LH=f'(u)[R(g)-u]\sqg+f[u]\sqg .\en{C.2'} $$
The corresponding field equations are nothing but the system
$(C.3)$, upon the substitution $p\rightarrow f'(u)$,
$r(p)\rightarrow u$. For quadratic \lagrangian s, which most
frequently occur in the literature, the two fields $u$ and $p$
coincide up to a constant factor. More generally, since the
$R$--regularity is still necessary to ensure the equivalence of
$(C.2')$ and (1.1), the choice of either $p$ or $u$ to represent the
independent scalar field occurring in the second--order picture
is, from the mathematical viewpoint, a mere matter of
convenience. The use of the scalar field $u$ allows one to bypass
the problem of finding explicitly the function $r(p)$, but then
the gravitational equation contains a rather complicated
dynamical term for the $u$--field,
$\nabla_{\mu}\nabla_{\nu}f'(u)-g_{\mu\nu}\dal f'(u)$, which
depends on the particular choice of $f(R)$ in (1.1). On the other
hand, there is a {\it physical} motivation to formulate the theory
in terms the scalar field $\phi$, which can be equivalently
defined either by $\phi\propto\ln p$ or by $\phi\propto \ln
f'(u)$, since this field interacts with the Einstein--frame metric in
the standard way (to this purpose, the problem of inverting $f'$
cannot be avoided). In the sequel of this Appendix, however, it
will be  convenient to write all the equations in terms of the
conjugate momentum $p$.

\subsection{C.2 Inverse Legendre Transformation.}

So far, we have recalled how a nonlinear \lagrangian\ can be
recast into a \HL. Now we present the inverse procedure: how
to find a nonlinear \lagrangian\ equivalent to a given \HL. In
fact, let $$ \LH(p;g,\Psi)=\sqg[p R - H(p;g,\Psi)] \en{C.5} $$ be
a generic \HL. The function $H$ plays the role of a
Hamiltonian\foot{This ``Hamiltonian'' has actually nothing to
do with the notion of energy and the ADM formalism in General
Relativity.}\ for a system of matter (denoted collectively by
$\Psi$) and gravity $g_{\mu\nu}$, thus it does not depend on
derivatives of the gravitational momentum $p$, while it
depends on covariant derivatives of $\Psi$ up to some order
(the semicolon separates the field variables which are
accompanied by their derivatives from those which are not). The
variation of the action $S_{_H}=\int\LH d^4x$ with respect to
$\Psi$ yields the equations of motion for matter, $$
\vder{}{\Psi}(\sqg H)=0\ ,\en{C.6} $$ while variation of it with
respect to $p$ yields an {\it algebraic\/} equation for the
canonical momentum, $$ R(g)-\pder{H(p;g,\Psi)}{p}=0\
.\en{C.7} $$ Any solution of this ``equation of motion'' for $p$ is
denoted by $P(R;g,\Psi)$. Finally, varying the metric one gets
(second--order) gravitational field equations $$
{1\over\sqg}\vder{}{g^{\mu\nu}}(\sqg pR)\equiv
pG_{\mu\nu}(g) -
\nabla_{\mu}\nabla_{\nu}p+g_{\mu\nu}\dal p
={1\over\sqg}\vder{}{g^{\mu\nu}}(\sqg H)\ ; \en{C.8} $$
clearly $(C.7)$ and $(C.8)$ are a generalization of $(C.3)$. One
now defines, with the aid of a solution to $(C.7)$, a nonlinear
\lagrangian\ for the metric and the matter: $$
\NL=P(R;g,\Psi)R\sqg-H[P(R;g,\Psi);g,\Psi]\sqg\ ,\en{C.9} $$
where $R$ stands for $R(g)$ (and is not and independent
variable).  To show the equivalence of the two \lagrangian s,
one finds the field equations resulting from the stationarity of
the action $$ \SNL = \int L_{_NL}d^4x = \int
d^4x\sqg[P(R;g,\Psi)R-W(g,\Psi)]\ ,\en{C.10} $$ where
$W(g,\Psi)\equiv H[P(R;g,\Psi);g,\Psi]$. For a variation
$\delta\Psi$ of the matter fields, one has  $$ \eqalign{
\delta\SNL=& \int d^4x \left\lbrace  R \left[\pder{(\sqg
P)}{\Psi}\delta\Psi +\sum_{n=1}\pder{(\sqg
P)}{\Psi_{,\mu_1\cdots\mu_n}}\delta\Psi_{,\mu_1\cdots\mu
_n} \right] -\left.\vder{(\sqg H)}{\Psi}\right|_{p=P}
\delta\Psi \right.+\cr &-\left. \left.\pder{(\sqg
H)}{p}\right|_{p=P}\pder{P}{\Psi}\delta\Psi
-\sum_{n=1}\left.\pder{(\sqg H)}{p}\right|_{p=P}
\pder{P}{\Psi_{,\mu_1\cdots\mu_n}}\delta\Psi_{,\mu_1
\cdots\mu_n} \right\rbrace =\cr  =& \int
d^4x\left\lbrace\sqg \left(R-
\left.\pder{H}{p}\right|_{p=P}\right)\left(\pder{P}{\Psi}
\delta\Psi +\sum_{n=1}\pder{P}{\Psi_{,\mu_1\cdots\mu_n}}
\delta\Psi_{,\mu_1\cdots\mu_n}\right) \right.\cr &-
\left.\left.\vder{(\sqg
H)}{\Psi}\right|_{p=P}\delta\Psi\right\rbrace = -\int
d^4x\left.\vder{(\sqg H)}{\Psi}\right|_{p=P}\delta\Psi=0\ ,}
$$ \ssu{hence the equations of motion for matter are} $$
\left.\vder{(\sqg H)}{\Psi}\right|_{p=P}=0\en{C.11} $$ and
are equivalent to $(C.6)$, since $(C.7)$ holds identically. The
latter plays also a crucial role in deriving field equations for the
metric. In fact, a variation $\delta g^{\mu\nu}$ in the action
yields $$ \delta\SNL=\int d^4x[P\delta(\sqg R)+\sqg R\delta
P - \delta(\sqg W)] $$ and $$ \sqg R\delta P - \delta(\sqg W)
=\sqg \left(R - \left.\pder{H}{p}\right|_{p=P}\right)\delta P
-\left.\vder{H}{g^{\mu\nu}}\right|_{p=P}\delta g^{\mu\nu}
$$ so that the equations for gravity are $$
P(R;g,\Psi)G_{\mu\nu}(g) -
\nabla_{\mu}\nabla_{\nu}P+g_{\mu\nu}\dal P
-
{1\over\sqg}\left.\vder{H}{g^{\mu\nu}}\right|_{p=P}=0\en{
C.12} $$ and these are equivalent to $(C.8)$. This shows that the
\lagrangian s $\LH$ and $\NL$ are equivalent, at least
whenever equation $(C.7)$ has a unique solution (see Appendix
E).

One now uses the general formalism described above to find out
a nonlinear \lagrangian, and the corresponding fourth--order
field equations in MJCF in Case II of Section 2, i.e.~when matter
is minimally coupled to the rescaled metric $\g_{\mu\nu}$.
The inverse transformation from the Einstein frame does not
lead back to the original VJCF but to another (conformally
related) spacetime metric. This makes no surprise since, after
adding interaction with matter in the Einstein frame, the inverse
transformation is applied to a different system. Let us see it in
more detail.

\subsection{C.3 Adding matter interaction in the Einstein
frame.}

The starting point is a nonlinear vacuum \lagrangian\ for pure
gravity. Denoting the metric field in this VJCF by
$\hg_{\mu\nu}$ one has $\Lvac=f(\hR)\hsqg$. One makes
the standard transformation to the Einstein frame, $p=f'(\hR)$
and $\g_{\mu\nu}=p\hg_{\mu\nu}$. Solving the latter
equation for $\hR$ one finds $\hR=r(p)$ and this defines the
direct and inverse Legendre maps, $p=\tder{f(r)}{r}$ and
$r=r(p)$. The field $p$, in the ``phase space'' of the system,
becomes an independent dynamical variable and the relation
$p=f'(\hR)$ is a consequence of the equations of motion
generated by the \HL, or equivalently (after conformal
rescaling) by the Einstein--frame \lagrangian\ (1.5), $$ \L_{\rm
vac}=\nsqg\left[\R-{3\over
2p^{2}}\g^{\mu\nu}p_{,\mu}p_{,\nu}-2V(p)\right] \en{C.13}
$$ \su{with} $$ 2V(p)={r(p)\over p} -{f[r(p)]\over p^{2}}\
.\en{C.14} $$ Assuming that the rescaled metric $\g_{\mu\nu}$
is physical, one now adds the minimal--coupling term
$2\lmat(\g;\Psi)$ to the \lagrangian\ $(C.13)$. This clearly
affects the equations of motion and therefore the relation
$p=f'(\hR)$, holding in vacuum, fails to be true in the presence
of matter.

The present formalism restricts the allowed matter \lagrangian
s to those which do not depend on derivatives of the metric.
Although stringent in itself, the restriction admits the physically
important cases of perfect fluid, pure radiation, fields of spin 0
and usual fields of spin 1. The full \lagrangian\ is then $$
\L=\nsqg\left[\R-{3\over
2p^2}\g^{\mu\nu}p_{,\mu}p_{,\nu}-
2V(p)+2\lmat(\g;\Psi)\right] \en{C.15} $$ The usual conformal
rescaling $\g_{\mu\nu}=p g_{\mu\nu}$ transforms $(C.15)$
into the following \lagrangian\ for $g_{\mu\nu}$, $p$ and the
matter: $$ \LH(p;g,\Psi)=\sqg\lbrace p R - 2p^2[V(p)-
\lmat(pg;\Psi)]\rbrace\ . \en{C.16} $$ It has the form of a
\HL\ $(C.5)$ because the transformation cancelled the kinetic
term for the scalar. The information about the original vacuum
\lagrangian\ $\Lvac$ is encoded in the potential $V(p)$. The
requirement that $\LH$ does not depend on derivatives of $p$
is met provided $\lmat$ contains no covariant derivatives (thus,
it may contain at most first derivatives of $\Psi$). The
Hamiltonian in the present case is $$ H(p;g,\Psi)= p\cdot r(p)-
f[r(p)]-2p^2\lmat(pg;\Psi) \en{C.17} $$ and the equation for
$p$ becomes (2.16), $R(g)-
r(p)+g^{\mu\nu}T_{\mu\nu}(\Psi,pg)=0$ ; the matter stress--
energy tensor is defined, as always, in terms of the physical
metric $\g_{\mu\nu}$ and expressed in terms of
$g_{\mu\nu}$. If $\lmat=0$ or $T_{\mu\nu}$ is traceless
(e.g.~matter is conformally invariant), then $R(g)=r(p)$ and
$p=f'(R)$ will be a solution; otherwise the solution
$P(R;g,\Psi)\ne f'(R)$.   The \HL\ $(C.12)$ corresponds to a
nonlinear \lagrangian\ $(C.9)$ for $g_{\mu\nu}$ and $\Psi$.
The NLG \lagrangian\ $\NL$ so obtained reduces to the
original vacuum one if the matter fields are set to vanish;
however, in general it is not equal to the sum of $\Lvac$ and a
matter term analogous to $\lmat$ (see the examples below). In
the purely metric picture, the relation between the two metrics is
expressed by   $\g_{\mu\nu}=P(R;g,\Psi)g_{\mu\nu}$, which
explicitly depends on matter. For this reason the MJCF metric
corresponding to a given physical metric $\g_{\mu\nu}$ does
not coincide in general with the VJCF metric $\hg_{\mu\nu}$,
which is implicitly defined by
$\g_{\mu\nu}=f'(\hR)\hg_{\mu\nu}$. In MJCF the equations
of motion for matter, $(C.11)$, read $$ \vder{}{\Psi}\Big[\sqg
p^2\lmat(pg;\Psi)\Big]_{p=P}=0 \en{C.18} $$ The fourth-order
field equations for gravity, $(C.12)$, can be put in a number of
equivalent forms. Here we recast them in the form which is
closest to that for the vacuum case. Upon using
$r(p)=R(g)+g^{\mu\nu}T_{\mu\nu}$ one finds that
$PG_{\mu\nu}+\half Pg_{\mu\nu}r(P)=P R_{\mu\nu}+\half
Pg_{\mu\nu}g^{\alpha\beta}T_{\alpha\beta}$, then defining a
scalar $M(R;g,\Psi)$ as the matter contribution to $P$,
$P=f'(R)+M(R;g,\Psi)$, one arrives at the following gravitational
field equations: $$\eqalignno{ Q_{\mu\nu}+
M(R;g,\Psi)R_{\mu\nu}-\nabla_{\mu}\nabla_{\nu}M+
&g_{\mu\nu}\dal M-\half g_{\mu\nu}\lbrace f[r(p)]-
f(R)\rbrace = \cr &= P(R;g,\Psi)[T_{\mu\nu}(Pg;\Psi)-\half
g_{\mu\nu}g^{\alpha\beta}T_{\alpha\beta}]\ .&(C.19)} $$
Here $Q_{\mu\nu}\equiv f'(R)R_{\mu\nu}-\half
f(R)g_{\mu\nu}-\nabla_{\mu}\nabla_{\nu}f'(R)
+g_{\mu\nu}\dal f'(R)$ is the l.h.s.~of equation (1.2) and
satisfies the generalized Bianchi identity
$\nabla^{\nu}Q_{\mu\nu}=0$ (Appendix A). For traceless
matter, $(C.19)$ reduces to equation (2.13). One sees that matter
and gravitational variables are inextricably intertwined in
$(C.19)$, thus these equations do not provide conservation laws
similar to those in General Relativity. Formally, four matter
conservation laws arise upon taking the divergence of $(C.19)$,
then the purely gravitational part, $Q_{\mu\nu}$, disappears
and one is left with interaction terms. The resulting equations,
however, are too complicated to be of any practical use even in
the simples cases.

Below we give some examples of finding the nonlinear
\lagrangian\ for a given form $f(\hR)$ of the vacuum
\lagrangian\ and for dust or a scalar field + electromagnetic
field as the matter.

\subsection{Example 1: quadratic \lagrangian\ and charged
scalar field.}

Let the vacuum \lagrangian\ in VJCF be
$\Lvac=(a\hR^2+\hR)\hsqg$. Then the ``vacuum'' inverse
Legendre map is $r(p)={1\over 2a}(p-1)$ and $f[r(p)]={1\over
4a}(p^2-1)$. In Einstein--frame metric
$\g_{\mu\nu}=p\hg_{\mu\nu}$ one adds the interaction with
a massive complex--valued scalar field $\psi$ minimally
coupled to electromagnetic field, $$ \eqalignno{ L =& \left[\R-
{3\over 2p^2}\g^{\mu\nu}p_{,\mu}p_{,\nu}-{(p-1)^2\over
4ap^2}\right.\cr &\left.-
\g^{\mu\nu}D_{\mu}\psi(D_{\nu}\psi)^* - m^2\psi\psi^*-
{1\over 8\pi}F_{\alpha\mu}F_{\beta\nu}\g^{\mu\nu}
\g^{\alpha\beta}\right]\nsqg &(C.20)} $$  Here
$D_{\mu}\psi\equiv\partial_{\mu}\psi-ieA_{\mu}\psi$. After
conformal rescaling $\g_{\mu\nu}=p g_{\mu\nu}$ this
\lagrangian\ becomes (up to a full divergence) $$
\LH=\left[pR-{1\over 4a}(p-1)^2-
pg^{\mu\nu}D_{\mu}\psi(D_{\nu}\psi)^* -
m^2p^2\psi\psi^*-{1\over
8\pi}F_{\alpha\mu}F_{\beta\nu}g^{\mu\nu}g^{\alpha\beta}
\right]\sqg \en{C.21} $$  and the equation $(C.14)$ is solved to
yield $$   P(R,g;\psi)=2\left(4m^2\psi\psi^*+{1\over
a}\right)^{-1}\left[R- g^{\mu\nu}D_{\mu}\psi(D_{\nu}\psi)^*
+{1\over 2a}\right]\ . \en{C.22} $$  The nonlinear \lagrangian\
$(C.9)$ in MJCF, $$ \eqalignno{  \NL(g,\psi,F)  =&
\sqg\left\lbrace\left(4m^2\psi\psi^*+{1\over a}\right)^{-
1}\left[R- g^{\mu\nu}D_{\mu}\psi(D_{\nu}\psi)^* +{1\over
2a}\right]^2+\right.\cr &-\left.{1\over
8\pi}F_{\alpha\mu}F_{\beta\nu}g^{\mu\nu}g^{\alpha\beta}-
{1\over 4a}\right\rbrace\ , &(C.23)} $$ generates the following
fourth--order gravitational field equations: $$ \eqalignno{
P(R,g;\psi)&R_{\mu\nu}
-\nabla_{\mu}\nabla_{\nu}P+g_{\mu\nu}\dal P-
PD_{(\mu}\psi(D_{\nu)}\psi)^*\cr -& \half
g_{\mu\nu}\left\lbrace\left(4m^2\psi\psi^*+{1\over
a}\right)^{-1}\left[R-
g^{\alpha\beta}D_{\alpha}\psi(D_{\beta}\psi)^* +{1\over
2a}\right]^2 -{1\over 4a}\right\rbrace\cr -& {1\over
4\pi}(F_{\alpha\mu}F_{\beta\nu}g^{\alpha\beta}
-{1\over4}g_{\mu\nu}g^{\alpha\rho}g^{\beta\sigma}
F_{\alpha\beta}F_{\rho\sigma})=0\ . &(C.24)} $$

\subsection{Example 2: quadratic \lagrangian\ and dust.}

As previously we choose in VJCF
$\Lvac=(a\hR^2+\hR)\hsqg$, so the functions $r(p)$, $f[r(p)]$
and $V(p)$ are as in Example 1. In the Einstein frame one adds
pressureless dust with energy density $\nrho$, thus the full
\lagrangian\ is $$ \L = \nsqg\left[\R-{3\over
2p^2}\g^{\mu\nu}p_{,\mu}p_{,\nu}-{(p-1)^2\over
4ap^2}+2\nrho\right] \ . \en{C.25} $$ Upon conformal
rescaling one finds the \HL $$ \LH=\sqg\left[pR-{1\over
4a}(p-1)^2+2p^2\nrho\right]\en{C.26} $$ and accordingly the
Legendre map in the presence of matter is $$
P(R,\nrho)={2aR+1\over 2a\nrho+1}\ . \en{C.27} $$ Simple
manipulations yield the nonlinear \lagrangian\ in MJCF, $$
\NL = {1\over 4a}\left[(12a\nrho+1)\left({2aR+1\over
2a\nrho+1}\right)^2 -1\right]\sqg \en{C.28} $$ and the field
equations for the metric, $$ P(R,\nrho)R_{\mu\nu}
-\nabla_{\mu}\nabla_{\nu}P+g_{\mu\nu}\dal P - {1\over
8a}(P^2-1)g_{\mu\nu}= P^2\nrho(u_{\mu}u_{\nu}+\half
g_{\mu\nu})\en{C.29} $$ It is very difficult to extract from
these equations any nontrivial information about motion of dust
particles.

\subsection{Example 3: logarithmic \lagrangian\ and dust.}

Cases where the equation $(C.18)$ for $P$ can be explicitly
solved are quite exceptional. Besides the quadratic \lagrangian\
and a simple matter source, one of these is provided by a
logarithmic \lagrangian\ (an exponential one is not the case)
and dust.

In vacuum Jordan frame one has $\Lvac={1\over
a}\ln(1+a\hR)\hsqg$ (for simplicity, we consider only the
sector on which $(1+a\hR)>0$) and proceeding as in the
previous cases one finds $$ p={1\over ar+1}>0\ ,\quad
r(p)={1\over a}\left({1\over p}-1\right)\ ,\quad f[r(p)]=-
{1\over a}\ln p\ , $$ $$  \quad V(p)={1\over 2ap}\left[{1\over
p}(\ln p+1)-1\right]\ , $$ and $$ \LH=\sqg\left[pR-{1\over
a}(\ln p-p+1)+2p^2\nrho\right]\ .\en{C.31} $$ The
corresponding Legendre map is \def\Ra{\left(R+{1\over
a}\right)} $$
P(R)={1\over8\nrho}\left[\sqrt{\Ra^2+{16\nrho\over a}}-
\Ra\right] $$ (there is another solution $P(R)<0$ which is
discarded) and the resulting nonlinear \lagrangian\ is
$$\eqalignno{ \NL=
&{\sqg\over16\nrho}\left\lbrace\Ra\left[\sqrt{\Ra^2+
{16\nrho\over a}}-\Ra\right]+\right.\cr &\left.-{16\nrho\over
a}\ln\left[\sqrt{\Ra^2+{16\nrho\over a}}-\Ra\right]-
{8\nrho\over a}[2\ln(8\nrho)-1]\right\rbrace\ ; &(C.32)} $$
notice that $\NL={1\over a}\ln(1+aR)\sqg+2\nrho (1+aR)^{-
2}\sqg+O(\nrho^2)$.

\section{Appendix D: The inverse problem of nonlinear gravity}

By means of the \HL\ method and the conformal rescaling it is
possible to map a nonlinear vacuum gravity theory with
\lagrangian\ $\Lvac=\sqg f(R)$ to a dynamically equivalent
system consisting of Einstein gravity and a minimally coupled
nonlinear scalar field with a potential determined by the
function $f$. The inverse problem of nonlinear gravity consists
in making an  inverse transformation. Given a scalar field which
self--interacts via an arbitrary potential $\U(\phi)$, is it possible
to map the \lagrangian\ $$ \L(\g,\phi)=\nsqg\left[\R-
\g^{\mu\nu}\phi_{,\mu}\phi_{,\nu}-2\U(\phi)\right]
\en{D.1} $$ in the Einstein frame to an equivalent \lagrangian\
$\NL=\sqg f(R)$ in the VJCF? With one exception, the answer is
``yes'', but in most cases the nonlinear \lagrangian\ cannot be
expressed int terms of elementary functions (even if $\U(\phi)$
can be). The procedure is as presented in Appendix C. First one
redefines the scalar by setting $\phi=\sth\ln p$ and then makes
the conformal rescaling $\g_{\mu\nu}=p g_{\mu\nu}$ to
cancel the kinetic term for the field $p$. After discarding a
divergence term, $(D.1)$ becomes a \HL, $$ \LH=\sqg[pR-
2p^2U(p)]\ ,\en{D.2} $$ where $U(p)\equiv\U(\sth\ln p)$.
According to $(C.7)$, the inverse Legendre map $p=P(R)$ is a
solution of the algebraic (w.r.~to $p$) equation $$ R(g)-
\tder{}{p}[2p^2U(p)]=0\ ;\en{D.3} $$ then the function $f$ is
determined by the relation $f'(R)=P(R)$.

An alternative method consists in solving an ordinary
differential equation directly for $f$. Comparing $(D.2)$ with
$(C.16)$ one sees that $U(p)\equiv V(p)$, where $V$ is given in
$(C.14)$ with $p=\tder{f(r)}{r}$ and $r(p)$ its inverse function.
Then the equation $U(p)=V(p)$ becomes a differential
equation\foot{In [52], Teyssandier and Tourrenc present a
system of differential equations which is equivalent to $(D.4)$.}
for $f(r)$ $$ 2(f')^2U(f')=r f'-f \en{D.4} $$ Solving this equation,
however, is by no means easier than solving $(D.3)$. Moreover,
to ensure the consistency of the Legendre transformation, a
solution to $(D.4)$ should also meet the $R$--regularity
condition $\tder{^2f}{r^2}\ne 0$. Differentiating $(D.4)$ w.r.~to
$r$ and using $f''\ne 0$ one arrives at an equivalent equation,
which added to $(D.4)$ yields $$
2f'\left.\tder{U}{p}\right|_{p=f'} + 2U(f') - {f\over (f')^2}=0\
,\en{D.5} $$ thus $f$ as a function of $p$ is given by $$
f[r(p)]=2p^2\left[U(p)+p\tder{U}{p}\right]\ ,\en{D.6} $$ while
$r(p)$ is determined from the form of $V$, $$
r(p)=2pU(p)+{1\over p}f[r(p)]\ .\en{D.7} $$ To determine $f(r)$
without any integration one should compute $f[r(p)]$ from
$(D.6)$, then solve $(D.7)$ for $p(r)$ (which is the same as
solving $(D.3)$) and insert the solution back to $f[r(p)]$.

First we note that the method does not apply to the simplest
case of the massless linear field, $\U(\phi)\equiv 0$. In fact, in
this case $(D.4)$ is solved by $f=Cr$ and this solution is
excluded because it is not $R$--regular. Thus, the above--
mentioned exception is provided by the Einstein--frame
\lagrangian\ $ \L(\g,\phi)=\nsqg\left[\R-
\g^{\mu\nu}\phi_{,\mu}\phi_{,\nu}\right] $, for which the
$\phi$--field cannot be eliminated to yield an equivalent purely
metric NLG theory. This \lagrangian, on the other hand, can be
turned into a STG \lagrangian\ (2.18) by a suitable conformal
rescaling. A constant potential in $(D.1)$ is interpreted as a
cosmological constant, $\U(\phi)=\Lambda$. Then $(D.6)$ and
$(D.7)$ easily yield $f(R)={1\over 8\Lambda}R^2$. This result
is also obtained as a particular (``singular'') solution of $(D.4)$,
which in this case is a Clairaut equation (and has also a general
solution, which should be excluded as being linear).

For nonconstant potentials $\U(\phi)$ the solutions do exist,
but these are practically inaccessible since one is unable to solve
$(D.3)$ or $(D.7)$. For the most interesting -- from the
field--theoretical viewpoint -- potential,
$\U(\phi)=\lambda\phi^n$, $\lambda={\rm const.}$,
$n=2,3,4$, one finds $ U(p)=\mu(\ln p)^n$, with
$\mu\equiv\left({3\over2}\right)^{n\over2}\lambda $, $$
f[r(p)]=2\mu p^2(\ln p+n)(\ln p)^{n-1}\en{D.8} $$ \su{and} $$
r(p)=2\mu p(2\ln p+n)(\ln p)^{n-1}\ .\en{D.9} $$

The Liouville field theory [32] provides one of the few examples
where equation $(D.7)$ can be explicitly solved. In this case
$\U(\phi)=A\exp(\alpha\phi)$, ($A$, $\alpha$ constants),
and  one finds $f[r(p)]=2(1+\beta)Ap^{\beta+2}$, with
$\beta\equiv\sth\alpha$; $r(p)=2(2+\beta)Ap^{\beta+1}$,
then  $$ P(R)=[2(2+\beta)A]^{^{-
{1\over\beta+1}}}R^{^{{1\over\beta+1}}}  $$ \su{and} $$
f(R)=2(1+\beta)A[2(2+\beta)A]^{^{{\beta+2\over\beta+1}}}
R^{^{{\beta+2\over\beta+1}}} \ . \en{D.10} $$

\section{Appendix E: Local dynamical equivalence with GR and
global solutions of NLG theories}

In Appendix C we have reviewed the Legendre--transformation
method. As in classical mechanics, the relationship between the
original fourth--order dynamics and the second--order
equations generated by the \HL\ is a ``dynamical equivalence''.
This means that there is a one--to--one correspondence between
{\it solutions\/} of the two systems of equations. This
correspondence is provided by the Legendre map. Since the
Legendre map is not, in general, globally invertible, the
correspondence (and thus the dynamical equivalence) holds
locally. A distinct problem is whether the conformal rescaling
leading from the Jordan frame to the Einstein frame can be
defined globally or not. In the case of quadratic \lagrangian s
only the second problem arises. If, for a given NLG lagrangian
(1.1), both the inverse of the Legendre map and the conformal
rescaling can be {\it globally\/} defined for all solutions of the
NLG equations (1.2), then the set of these solutions actually
coincides with the set of solutions of the Einstein--frame
equations (General Relativity + scalar field). Both conditions,
however, do not hold in general, and therefore one should make
a distinction between local solutions and global solutions. This
does not mean that the local dynamical equivalence with GR is
irrelevant while dealing with global solutions of NLG; on the
contrary, it allows to understand completely how global
solutions of NLG theory look like. In this Appendix we discuss
separately the problems arising when the Legendre map is not
bijective and when the conformal factor vanishes.

Once one is given a NLG lagrangian (1.1), the Legendre map is
defined, as it is explained in Appendix C, by the function
$p=f'(R)$. Whenever the Legendre map can be inverted, one
constructs the \HL\ and the corresponding second--order
equations. The dynamical equivalence of the latter with NLG
equations means, in one direction, that for {\it any\/} solution
of the second--order system $(C.3)$, represented in a local
coordinate system by
$(g_{\mu\nu}(x^{\alpha}),p(x^{\alpha}))$, the metric tensor
$g_{\mu\nu}(x^{\alpha})$ is a solution of the original
fourth--order NLG equations. Conversely, to any given solution
$g_{\mu\nu}(x^{\alpha})$ of the NLG equations, one can
associate the corresponding  curvature scalar
$R=R(x^{\alpha})$; one introduces the scalar field
$p(x^{\alpha})=f'[R(x^{\alpha})]$,  and the pair
$(g_{\mu\nu}(x^{\alpha}),p(x^{\alpha}))$  then fulfills the
system $(C.3)$. The notion of {\it dynamical equivalence\/}
means nothing more than that. In particular, the value of the
\HL\ for a given pair $(g,p)$ coincides with the value of the
NLG \lagrangian\ (1.1) for the same metric $g$ {\it only\/} if
$(g,p)$ is a {\it solution\/} of the field equations; two
dynamically equivalent action functionals have the same
stationary points, but they take different values at states of the
system which do not correspond to classical solutions.

The dynamical equivalence holds under the $R$--regularity
condition $f''(R)\ne 0$. For the sake of simplicity, we first
consider a vacuum NLG (1.1) with $f$ analytic. Then the
equation  $$ f''(R)= 0 \en{E.1} $$ is an algebraic equation which
has a discrete set of solutions $R=k_i$ (the label $i$ runs over a
finite set of integers if $f$ is polynomial). To these values of the
curvature scalar corresponds, under the Legendre map, a set of
values $p=c_i=p(k_i)$ of the scalar field (the correspondence
between the sets $\lbrace k_i\rbrace$ and $\lbrace c_i\rbrace$
is not bijective, in general). The equations of the system $(C.3)$
are not defined for $p=c_i$, because the domain of the inverse
Legendre map $r(p)$, which occurs explicitly in $(C.3)$, does
not include these points. A global dynamical equivalence holds
whenever one can ensure that, for {\it all\/} possible solutions
of the NLG equations, the curvature scalar nowhere attains one
of the values $k_i$ at any point of space--time. This is actually
possible only if $(D.1)$ has no real solutions at all: this happens,
e.g., for $f(R)=R+aR^2$, $e^{aR}$, $\cosh(aR)$,
$\sum_{k=1}^na_kR^{2k}$ with $a_k>0$.

If the set of critical curvatures $\lbrace k_i\rbrace$ is not empty,
the solutions of NLG equations (1.2) can be divided into three
groups. First, the NLG equations may admit constant curvature
solutions for which $R(x^{\alpha})\equiv k_i$ (for some $i$) in
the whole space--time. Such solutions do not correspond to any
solution of the second--order system $(C.3)$. However, in both
cases it is very easy to check by direct computation whether
such solutions exist, and take them properly into account. The
second group of solutions includes global solutions which do
not attain any critical value at any point; for each solution of this
type we can find a corresponding global solution of $(C.3)$. The
third possibility is that a (non--constant) solution of NLG
equations attains some of the critical values in some regions of
space--time. Under reasonable assumptions on the regularity of
the metric tensor, this region has measure zero: it is a
lower--dimensional submanifold of space--time. This critical
submanifold, which is different for each solution, may have
several disconnected components, corresponding to different
critical values. The critical submanifold separates space--time
into open domains on which the $R$-regularity condition is
fulfilled. Inside each regularity domain, one can associate to the
global NLG solution under consideration a local solution of the
system $(C.3)$. Such a local solution
$(g_{\mu\nu}(x^{\alpha}),p(x^{\alpha}))$ has the property
that $p(x^{\alpha})\rightarrow c_i$, for some $i$, when one
approaches the critical submanifold, thus it can be extended by
continuity across the critical surfaces. As a matter of fact, the
system $(C.3)$ itself can be extended to such points as well: the
derivative of $r(p)$ blows up at critical points, but $r(p)$ admits
a finite limit there.

The most relevant problem which can arise from a failure of the
regularity condition on lower--dimensional submanifolds is
connected with the uniqueness properties of solutions. A given
local solution may admit several continuations outside its
regularity domain. This is connected with (but not entirely
dependent on) the following fact, which we have not yet taken
into account: whenever $f'(R)$ is not bijective, then in each
regularity domain the inverse function may not be unique.

In general, the equation $f'(R)\big|_{R=r(p)}- p=0$ admits
several roots for a given value of $p$. Assume for instance that
$f'(r)$ is a polynomial of third degree; then, for a generic choice
of the coefficients there are two critical values $\lbrace
k_1,k_2\rbrace$ ($k_1<k_2$), and one finds one root $r_1(p)$
for $-\infty<p<k_1$, three distinct roots $\lbrace
r_i(p)\rbrace_{i=2,3,4}$ for $k_1<p<k_2$ and again one root
$r_5(p)$ for $p>k_2$. Putting together the families of roots
which have the same limit for $p\rightarrow k_i$, one obtains
exactly three inverse Legendre maps: $r_-(p)$, defined for
$p\in(-\infty,k_2)$; $r_o(p)$, defined only for $p\in(k_1,k_2)$;
and $r_+(p)$ for $p\in(k_1,\infty)$. Thus one has to introduce
three distinct \HL s, each one containing a different potential
for $p$. Each \HL\ is defined only for the range of values of
$p$ for which the potential is defined. In the example
considered above, for $p\in(k_1,k_2)$ there are three different
second--order scalar--tensor systems $(C.3)$ locally equivalent
to (1.2) (we may then speak of different ``sectors'' for the scalar
field). It might seem that this causes troubles such as ill-
posedness of the Cauchy problem, but it is not so.  In fact, the
ranges for $p$ may overlap (and they do, in general), but the
ranges of the curvature scalar $R$ corresponding to different
potential are always disjoint, since the Legendre map $p=p(R)$
is globally and uniquely defined. Thus, if one considers a global
NLG solution, in each regularity domain one knows exactly to
which sector the scalar field belongs, just by checking the range
of values of $R$, and therefore one knows without ambiguity
which scalar--tensor equations $(C.3)$ are fulfilled by the pair
$(g,p\equiv f'(R))$. If, in the opposite direction, one considers a
Cauchy problem for the scalar--tensor model, then the Cauchy
data are compatible with at most one sector of the theory, and
therefore determine the equations which have to be solved.

Hence, the {\it local dynamical equivalence\/} which holds in
the cases we are dealing with can be described as follows: to
each solution of the NLG equations (1.2), with the exception of
the solutions for which $f''(R)\equiv 0$, corresponds a
scalar--tensor pair $(g,p)$. This pair has one or more
$R$--regularity domains, which are the open regions of
space--time on which $f''(R)\ne 0$. Inside each regularity
domain, the pair $(g,p)$ coincides with a solution of the
scalar--tensor model described by a suitable \HL\ $(C.2)$,
where $r(p)$ is the {\it unique\/} inverse Legendre map
defined for the range of values of $R$ occurring in {\it that\/}
regularity domain. Conversely, {\it all\/} solutions of NLG
equations (besides the particular constant--curvature solutions
mentioned above) can be recovered by matching together local
solutions of the scalar--tensor models corresponding to all
possible inverse Legendre maps $r(p)$ (for the given NLG
\lagrangian); each local solution exists, by definition, only in the
open region of space time on which the scalar--tensor equation
$(C.3)$ is well-defined; however, in general the solution can be
extended to the closure of the domain of definition of the
equations, and can therefore be matched with analogous local
solutions on adjacent domains; the minimal glueing prescription
is that the curvature scalar of the metric tend to the same limit
on both sides of a critical surface. This entails the continuity of
$p$, but is a stronger assumption because $R$ could have a
discontinuity, due to the jump from one sector to another, even
when $p$ is continuous across the critical surface. The matching
requirement, therefore, prevents certain pairs of sectors to occur
in adjacent domains.

The existence of different sectors depends on the
non--uniqueness of the inverse of $f'(R)$. By a different
parametrization of the scalar--tensor model, namely by using
the alternative  \HL\ $(C.2')$, one avoids this problem. The
\HL, and therefore the scalar--tensor equations, need
nevertheless to be continued to critical surfaces (whereby the
equation (1.2) cannot be recast in normal form); the continuation
may be not regular enough to avoid well-posedness problems,
and the distinction between global and local dynamical
equivalence should be kept. Moreover, we have already
mentioned in Appendix C that the definition of the {\it
physical\/} scalar field $\phi$ requires in any case the use of an
inverse Legendre map; thus, the physical picture of the theory
does include different sectors, unless the NLG \lagrangian\ is
globally $R$-regular.

In Appendix C, we have also showed how a nonlinear
\lagrangian\ can be recovered, by inverse Legendre
transformation, after a modification of the \HL\ by the addition
of matter coupling terms; however, we did not take into account
the possible non--uniqueness of the \HL.  We now complete the
discussion on this point by checking that,  in the case of multiple
\HL s corresponding to the same vacuum NLG \lagrangian,
matter coupling terms can be consistently added in all sectors,
so that the inverse Legendre transformation leads to a {\it
unique\/} nonlinear interaction \lagrangian. More explicitly,
suppose that for a given $f(R)$ in the NLG \lagrangian (1.1)
there exist several inverse Legendre maps, $\lbrace
r_i(p)\rbrace $, defined on different ranges of the scalar field
$p$ (which partially overlap). Accordingly, one has several
locally equivalent \HL s, each one containing a different
``Hamiltonian'' (we follow the notation of Appendix C) $H_i(p)
= p\cdot r_i(p) - f[r_i(p)]$.   Now, what happens if one modifies
each \HL\ as described by equation $(C.16)$? One would
reasonably expect that matter couples to the metric in the same
way for all sectors, since in the Einstein frame matter is not
coupled at all with the scalar field $p$: the ``Hamiltonians''
should hence become $H_i(p) = p\cdot r_i(p) - f[r_i(p)] -
2p^2\lmat(pg;\psi)$. Do such ``Hamiltonians'' correspond to a
unique nonlinear \lagrangian? Using the explicit form of
$H_i(p)$, the equation $(C.7)$ becomes   $$ R(g) - r_i(p) +
4p\,\lmat(pg;\psi) +
2p^2\pder{\lmat}{\g_{\mu\nu}}g_{\mu\nu}=0 \en{E.2} $$ Let
us redefine the auxiliary scalar field by setting $p=f'(u)$, as
explained in Appendix C: one has in each sector $u=r_i(p)$,
i.e.~$u$ is a multivalued function of $p$; for $u$, however, the
following unique equation holds: $$ R(g) - u +
4f'(u)\,\lmat[f'(u)g;\psi] +
2[f'(u)]^2\pder{\lmat}{\g_{\mu\nu}}g_{\mu\nu}=0 \en{E.3}
$$ leading to solutions $u=U(R;g,\psi)$ which are independent
of $i$. For any such solution, $P(R;g,\psi)=f'[U(R;g,\psi)]$
solves $(D.2)$ for all $i$, wherever the domain of $r_i(p)$
intersects the range of $P(R;g,\psi)$; thus, we see that even for
ranges of values of $p$ for which several \HL s coexist, $(E.2)$
admits solutions which do not depend on $i$. A nonlinear
\lagrangian\ can then be defined  by setting $$ L_{_{NL}}=
P(R;g,\psi)[R(g)-
U(R;g,\psi)]\sqg+f(U)\sqg+2[f(U)]^2\lmat[f(U)g;\psi]\sqg $$
For a more detailed investigation of this topic, we refer the
reader to [53].

We now consider the second problem: namely, whether the
Einstein conformal frame corresponding to a solution of NLG
equations can be globally defined or not. If we assume, as we
did in Section 1, that the conformal factor coincides with the
scalar field $p=f'(R)$,  the Einstein frame is globally defined
provided $$  f'(R)>0 \en{E.4} $$  everywhere. Otherwise, the
rescaling introduces new singularities.  This problem has been
raised by several authors; most of them, however, consider the
procedure leading from the metric $g$  in (1.1) to the pair
$(\g,\phi)$ occurring in the Einstein--frame \lagrangian\ (1.6)
as a single transformation, while it actually consists of three
distinct steps (addition of a new independent variable,
conformal rescaling, redefinition of the scalar field
$\phi=\ln(p)$). For instance,  Maeda in [5] comments about the
local character of the equivalence between the two frames, but
he refers only to possible failures of condition $(E.4)$, although
the proof of the equivalence which he gives in the same paper is
valid only if $(E.1)$ is fulfilled; moreover, his comment refers
mainly to the problem of defining the scalar field $\phi$
(denoted by $\psi$ in his paper) at points where $f'(R)=0$,
rather than to the fact that the Einstein metric vanishes at these
points.

The transformation leading from the scalar--tensor equations
$(C.3)$ to the Einstein--frame equations (1.4), which is the only
genuine conformal rescaling occurring in this context,  is
mathematically a mere change of variables. In contrast to the
case of the dynamical equivalence relating (1.1) to the \HL, the
latter is now simply re--expressed in terms of the new variables,
so both the \HL\ and the Einstein--frame \lagrangian\ (1.5)
take the same values (more precisely, differ by a total
divergence) at conformally--related pairs $(g,p)$ and $(\g,p)$,
even if these pairs are {\it not\/} solutions of the field equations
(this fact is mentioned by Wands [6] for the case of conformal
equivalence between JBD theory and GR+scalar field).  The
condition $(E.4)$ is, in principle, completely independent from
$(E.1)$. Only for particular NLG \lagrangian s, e.g. $f(R)=R^k$,
with $k>2$, $p=0$ is also a critical value for the Legendre map,
and the two conditions partially overlap.

In some cases, we can ensure from the very beginning that $p$
is positive everywhere. If $f(R)$ is a polynomial of odd degree
with positive highest--order coefficient, then the Legendre map
$p(R)$ is bounded fom below, and the coefficients of the
lower--order terms in $f(R)$ can be chosen so that $p\ge p_{\rm
min}>0$.

If $f(R)$ is a polynomial of even degree (for instance, in the
quadratic case), then the scalar field $p$ will range over the full
real line. In this case, we should again regard the solutions of
NLG as divided into three groups. For some solutions we may
have $p\equiv 0$; such solutions have constant scalar curvature
and can be easily singled out and treated separately (under the
assumption (3.8), flat space dose {\it not\/} belong to this class
of solutions). For other solutions, $(E.4)$ holds everywhere in
space--time and the Einstein frame is then globally defined. The
problem arises for the third group of solutions, for which $p$
vanishes on some region of space--time. As previously, under
suitable assumptions such regions have zero measure and in
most cases are lower--dimensional submanifolds of space--time.
For these solutions, the conformal transformation introduces
singularity surfaces. In rigorous mathematical terms we should
say that the equivalence holds only where both $g_{\mu\nu}$
and $\g_{\mu\nu}$ are nondegenerate; both the \lagrangian\
and the field equations for $\g_{\mu\nu}$ are defined only
where  $\g_{\mu\nu}$ is regular. However, for some
\lagrangian s the Einstein--frame equations admit a solution
$(\g,p)$ such that $\g_{\mu\nu}$ is nondegenerate but $p$
vanishes, and therefore it is the Jordan--frame metric which
cannot be recovered at such points: whether we should regard
this situation as ``singular'' or not, from the physical viewpoint,
depends on the physical significance attached to the metric.

A separate problem, which has an evident physical relevance,
concerns the signature of the rescaled metric. Whenever $p<0$,
the conformal rescaling can be defined but it changes the
signature of the metric. If we assume that the Einstein--frame
metric is the physical one, this is not acceptable because the
choice of the signature, although arbitrary, should be globally
consistent. This problem is solved by some authors (e.g. Barrow
\& Cotsakis and Maeda [5]) by changing the definition of the
Einstein metric to $$ \g_{\mu\nu} = |p| g_{\mu\nu} .\en{E.5}
$$ This introduces a discontinuity in the first derivatives of the
metric, but this happens exactly on the singularity surfaces, and
constitutes a minor problem. The definition, however, causes an
overall minus sign to appear in front of the Einstein--frame
\lagrangian\ when $p$ is negative (in other words, the entire
\lagrangian\ is multiplied by ${p\over|p|}$; this seems to
make the \lagrangian\ discontinuous at $p=0$, but in fact the
\lagrangian\ itself is not defined at all there). This sign has to be
taken into account while adding a matter \lagrangian, since the
sign of the latter should agree with the sign of the
Einstein--Hilbert term $\R\nsqg$.

An alternative viewpoint is to reject the requirement that both
metrics have the same signature. Since we claim that only one
metric is physical, any requirement on the signature should
affect only this one. Once we have found a physically acceptable
solution $(\g,p)$, the possible fact that the corresponding
Jordan--frame metric changes signature due to a change of sign
of $p$ may be regarded as irrelevant (anyhow, to change
signature, the Jordan frame metric should necessarily become
singular somewere, and this feature is much more relevant).
\vfill\eject

\frenchspacing \parindent=40pt \parskip=4pt {\bf References}

\def\ref[#1]{\medskip\hang\indent\llap{[#1] --
}\ignorespaces}

\def\NC{{\it Nuovo Cim.\ }} \def\JP{{\it J. Phys.\ }}
\def\PL{{\it Phys. Lett.\ }} \def\PR{{\it Phys. Rev.\ }}
\def\PRL{{\it Phys. Rev. Lett.\ }} \def\NP{{\it Nucl. Phys.\ }}
\def\CQG{{\it Class. Quantum Grav.\ }} \def\GRG{{\it Gen.
Rel. Grav.\ }} \def\JMP{{\it J. Math. Phys.\ }} \def\IJMP{{\it
Intern. J. Mod. Phys.\ }} \def\PTP{{\it Prog. Theor. Phys.\ }}
\def\PRSL{{\it Proc. Roy. Soc. London\ }} \def\APNY{{\it
Ann. Phys. (NY)\ }} \def\CMP{{\it Commun. Math. Phys.\ }}

\ref[1] T. Damour and G. Esposito--Far\`ese, \CQG 9 (1992)
2093

\ref[2]  P.W. Higgs, \NC 11 (1959) 816

G. Bicknell, \JP A7 (1974) 1061

B. Whitt, \PL 145B (1984) 176

\ref[3] G. Magnano, M. Ferraris and M. Francaviglia, \GRG 19
(1987) 465

\ref[4] A. Jakubiec and J. Kijowski, \GRG 19 (1987) 719

A. Jakubiec and J. Kijowski, \PR D39 (1989) 1406

A. Jakubiec and J. Kijowski, \JMP 30  (1989)  1073

\ref[5] J.D. Barrow and S. Cotsakis, \PL B214 (1988) 515

K. Maeda, \PR D39 (1989) 3159

S. Gottl\"ober, H.--J. Schmidt and A.A. Starobinsky, \CQG 7
(1990) 893

H.--J. Schmidt, \CQG 7 (1990) 1023

\ref[6] D. Wands, preprint Sussex--AST 93/7--1

\ref[7] K. Maeda, \PR D36 (1988) 858

J.D. Barrow and K. Maeda, \NP B341 (1990) 294

A. Berkin, K. Maeda and J. Yokoyama, \PRL 65 (1990) 141

T. Damour, G. Gibbons and C. Gundlach, \PRL 64 (1990) 123

T. Damour and C. Gundlach, \PR D43 (1991) 3873

R. Holman, E.W. Kolb and Y. Wang, \PRL 65 (1990) 17

R. Holman, E.W. Kolb, S. Vadas and Y. Wang, \PR D43 (1991)
995

R. Fakir and W.G. Unruh, \PR D41 (1990) 1783

S. Kalara, N. Kaloper and K. Olive, \NP B341  (1990) 252

S. Mollerach and S. Matarrese, \PR D45 (1992) 1961

G. Piccinelli, F. Lucchin and S. Matarrese, \PL B277 (1992) 58

Z.--J. Tao and X. Xue, \PR D45 (1992) 1878

A. Wu, \PR D45 (1992) 2653

S. del Campo, \PR D45 (1992) 3386

I. Tkachev, \PR D45 (1992) R4367

S. Mignemi and D. Wiltshire, \PR D46 (1992) 5329

J.D. Barrow, \PR D47 (1993) 1475

S. Cotsakis and G. Flessas, \PR D48 (1993) 3577

A. Laycock and A. Liddle, preprint Sussex--AST 93/6--1

\ref[8] T. Damour and K. Nordtvedt, \PR D48 (1993) 3436

\ref[9] W. Bruckman and E. Velazquez, \GRG 25 (1993) 901

\ref[10] G.W. Gibbons and K. Maeda, \NP B298 (1988) 741

\ref[11] L. Pimentel and J. Stein--Schabes, \PL B216 (1989) 27

J. Alonso, F. Barbero, J. Julve and A. Tiemblo, preprint IMAFF
93/9

L. Amendola, D. Bellisai and F. Occhionero, \PR D47  (1993)
4267

\ref[12] D. Salopek, J. Bond and J. Bardeen, \PR D40 (1989) 1753

E.W. Kolb, D. Salopek and M.S. Turner, \PR D42 (1990) 3925

\ref[13] Y. Fujii and T. Nishioka, \PR D42 (1990) 361

T. Nishioka and Y. Fujii, \PR D45 (1992) 2140

\ref[14] Y.M. Cho, \PRL 68 (1992) 3133

\ref[15] Y. Kubyshin, V. Rubakov and I. Tkachev, \IJMP A4
(1989) 1409

N. Deruelle, J. Garriga and E. Verdaguer, \PR D43 (1991) 1032

\ref[16] W. Buchm\"uller and N. Dragon, \NP B321 (1989) 207

N. Makino and M. Sasaki, \PTP 86 (1991) 103

R. Holman, E.W. Kolb, S. Vadas, Y. Wang and E. Weinberg, \PL
B237 (1990) 37

B.A. Campbell, A.D. Linde and K. Olive, \NP B355 (1991) 146

J.A. Casas, J. Garcia--Bellido and M. Quiros, \NP B361 (1991)
713

J.A. Casas, J. Garcia--Bellido and M. Quiros, \CQG 9 (1992) 1371

L. Garay and J. Garcia--Bellido, \NP B400 (1993) 416

\ref[17] N. Deruelle and P. Spindel, \CQG 7 (1990) 1599

J.--C. Hwang, \CQG 7 (1990) 1613

S. Gottl\"ober, V. M\"uller and A.A. Starobinsky, \PR D43
(1991) 2510

Y. Suzuki and M. Yoshimura, \PR D43 (1991) 2549

T. Rothman and P. Anninos, \PR D44 (1991) 3087

E. Guendelman, \PL B279 (1992) 254

A. Guth and B. Jain, \PR D45 (1992) 426

A. Liddle and D. Wands, \PR D45 (1992) 2665

\ref[18] C.H. Brans, \CQG 5 (1988) L197

\ref[19] M. Ferraris, M. Francaviglia and G. Magnano, \CQG 7
(1990) 261

L.M. Soko\l owski, \CQG 6 (1989) 2045

\ref[20] A. Strominger, \PR D30 (1984) 2257

\ref[21] S. Cotsakis, \PR D47 (1993) 1437

\ref[22] K.S. Stelle, \PR D16 (1977) 953

K.S. Stelle, \GRG 9 (1978) 353

\ref[23] N. Barth and S.M. Christensen, \PR D28 (1983) 1876

\ref[24] T. Singh and T. Singh, \IJMP A2 (1987) 645

\ref[25] C.M. Will, ``Theory and Experiment in Gravitational
Physics'' Cambridge Univ. Press Cambridge 1981 \S 5

\ref[26] R.H. Dicke, \PR 125 (1962) 2163

\ref[27] R. Penrose, \PRSL A284 (1965) 159

C.G. Callan Jr., S. Coleman and R. Jackiw, \APNY 59 (1970) 42

L. Parker, \PR D7 (1973) 976

\ref[28] J.D. Bekenstein, \APNY 82 (1974) 535

\ref[29] C. Klim\v c\'\i k and P. Kolnik, \PR D48 (1993) 616

\ref[30] M.S. Madsen, \CQG 5 (1988) 627

\ref[31] E. Witten, \NP B195 (1982) 481

\ref[32] E. d'Hoker and R. Jackiw, \PR D26 (1982) 3517

\ref[33] L.F. Abbott and S. Deser, \NP B195 (1982) 76

\ref[34] R. Streater and A. Wightman, ``PCT, Spin and Statistics,
and All That'' Benjamin New York 1964

H. Epstein, V. Glaser and A. Jaffe, \NC 36 (1965) 1016

\ref[35] L. Ford and T. Roman, \PR D46 (1992) 1328

L. Ford and T. Roman, \PR D48 (1993) 776

\ref[36] S.W. Hawking and G.F.R. Ellis, ``The Large Scale
Structure of Space--time'' Cambridge Univ. Press Cambridge
1973

H. Kandrup, \PR D46 (1992) 5360

\ref[37] T.T. Wu and C.N. Yang, \PR D13  (1976) 3233

\ref[38] M.J. Duff, in: ``Quantum Gravity 2, A Second Oxford
Symposium'' ed. C. Isham, R. Penrose and D. Sciama Clarendon
Oxford 1981 p. 81

\ref[39] I.J. Muzinich and M. Soldate, \GRG 21 (1989) 307

\ref[40] G.T. Horowitz, ``The Positive Energy Theorem and its
Extensions'' in: ``Asymptotic Behavior of Mass and Spacetime
Geometry'' Lect. Notes in Physics 202, ed. F. Flaherty Springer
Berlin 1984 p. 1

\ref[41] G. Magnano, M. Ferraris and M. Francaviglia, \CQG 7
(1990) 557

\ref[42] H. Poincar\'e, ``M\'ethodes nouvelles de la
M\'ecanique C\'eleste''  vol. III, \S 29 Paris 1899

T. Levi--Civita and U. Amaldi, ``Lezioni di Meccanica
Razionale'' vol. II, \S 11, par. 7 Bologna 1927

\ref[43] L.M. Soko\l owski, Z. Golda, M. Litterio and L.
Amendola, \IJMP A6 (1991) 4517

\ref[44] E. Witten, \CMP 80 (1981) 381

O. Reula, \JMP 23 (1982) 810

T. Parker and C. Taubes, \CMP 84 (1982) 223

\ref[45] D. Brill and H. Pfister, \PL B228 (1989) 359

\ref[46] D. Brill and G.T. Horowitz, \PL B262 (1991) 437

\ref[47] S. Deser and Z. Yang, \CQG 6 (1989) L83

\ref[48] D.G. Boulware, S. Deser and K.S. Stelle, \PL 168B (1986)
336

\ref[49] J.D. Bekenstein \PR D5 (1972) 1239

L.M. Soko\l owski and B. Carr \PL 176B (1986) 334

\ref[50] G.T. Horowitz ``String theory as a quantum theory of
gravity'' in: ``GRG 1989, Proc. of the 12th Conference on GRG,
Boulder, July 1989'', ed. N. Ashby, D. Bartlett and W. Wyss,
Cambridge Univ. Press Cambridge 1990 p. 419

D. Garfinkle, G.T. Horowitz and A. Strominger \PR D43 (1991)
3140

\ref[51] G. Magnano, M. Ferraris and M. Francaviglia, \JMP 31
(1990) 378

\ref[52] P. Teyssandier and P. Tourrenc, \JMP 24  (1983) 2793

\ref[53] M. Francaviglia, G. Magnano and I. Volovich, in
preparation

\vskip0pt plus 4fill \parindent=0pt  {\lineskiplimit=8pt
\baselineskip=9pt \newdimen\sccm \newdimen\scpt
\sccm=0.66truecm \scpt=0.66truept \font\rm=cmr10 scaled 700
\font\bf=cmb10 scaled 800 \font\it=cmti10 scaled 700
\textfont0=\scriptfont0 \textfont1=\scriptfont1
\textfont2=\scriptfont2 \textfont3=\scriptfont3
\textfont4=\scriptfont4 \textfont5=\scriptfont5
\textfont6=\scriptfont6 \hfuzz=5pt \hbadness=10000
\vfuzz=5pt

\rm

\def\vtrat#1{\kern-1.5pt\vbox{\leaders\vbox
to2\scpt{\vfill\hbox
to3pt{$\cdot$}\vfill}\vskip#1\sccm}\kern-1.5pt}
\def\htrat#1{\hbox{\leaders\hbox
to2\scpt{\hss$\cdot$\hskip-1\scpt}\hskip#1\sccm}}
\def\hltrat#1{\hbox{\leaders\hbox to2\scpt{\hss.\hskip-
1\scpt}\hskip#1\sccm}} \def\piu{{\rm +\ }}

\long\def\block#1#2#3#4{\vbox{ \hbox
to#1\sccm{\hfill#3\hfill} \hbox{\vrule\vbox to #2\sccm{
\hrule\vfill\hbox
to#1\sccm{\hfill#4\hfill}\vfill\hrule}\vrule}}}

\long\def\mblock#1#2#3#4{\vbox{ \hbox
to#1\sccm{\hfill#3\hfill}\vskip3\scpt \hbox{\vtrat{#2}\vbox
to #2\sccm{ \htrat{#1}\vfill\hbox
to#1\sccm{\hfill#4\hfill}\vfill\hltrat{#1}}\vtrat{#2}}}}

 \def\freccia{\vbox to1.7\sccm{\vfill\hbox to
0.5\sccm{\leaders\hrule\hfill}\vfill}} \def\pfreccia{\vbox
to1.7\sccm{\vfill\hbox to
0.5\sccm{\leaders\hrule\hskip0.65\sccm}\vfill}}

 \setbox1=\hbox{\hsize=3\sccm\leftskip=0\scpt
plus1fill\rightskip=0\scpt plus1fill\vbox{ without\vskip-
2\scpt cosmological\vskip-3\scpt function\vskip-2\scpt
$\lambda(\varphi)\equiv0$}}

\setbox2=\hbox{\hsize=4.9\sccm\leftskip=0\scpt
plus1fill\rightskip=0\scpt plus1fill\vbox{
$\omega(\varphi)\equiv0$\break {\it (Helmholtz Lagr.,
$\varphi\equiv p$)}}}

\setbox3=\hbox{\hsize=5\sccm\block{5.4}{1.1}{\vbox{\vfil
with cosmological function\vfil
\hbox{\hskip1.7\sccm$\lambda(\varphi)\not\equiv0$}\vskip
2\scpt}}{\vbox{\box2}}}

\setbox4=\hbox{\hsize=5 \sccm\leftskip=0\scpt
plus1fill\rightskip=0\scpt plus1fill\vbox{ {\bf Dynamical
equivalence} {\it (Legendre transformation)} regular at $R=0$ iff
$a\ne 0$ }}

\setbox5=\vbox{\block{6}{1}{}{nonlinear potential $V(\phi)$}
\block{6}{1}{}{$m=0\ ,\quad\Lambda\ne0$}
\block{6}{1}{}{$m=0\ ,\quad\Lambda=0$}}

\setbox6=\hbox{\hsize=5 \sccm\leftskip=0\scpt
plus1fill\rightskip=0\scpt plus1fill\vbox{ {\bf Dynamical
equivalence} {\it (Legendre transformation \break\piu
conformal rescaling \break\piu redef.~of the scalar field)}
regular at $R=0$ iff $\kappa\ne 0$ and $a\ne0$ }}

\setbox7=\hbox{\hsize=5 \sccm\leftskip=0\scpt
plus1fill\rightskip=0\scpt plus1fill\vbox{ {\bf Change of
variables} {\it (conformal rescaling \piu redef.~of the scalar
field)} regular for $\varphi\ne0$ }}

\setbox8=\hbox{\hsize=5 \sccm\leftskip=0\scpt
plus1fill\rightskip=0\scpt plus1fill\vbox{ {\bf Change of
variables} {\it (conformal rescaling \piu redef.~of the scalar
field)} regular for $\varphi\ne0$ }}

\setbox9=\hbox{\hsize=5 \sccm\leftskip=0\scpt
plus1fill\rightskip=0\scpt plus1fill\vbox{ {\bf Change of
variables} {\it (Bekenstein transf. = conf. rescaling \piu
redef.~of the scalar)} }}

\setbox10=\hbox{\hsize=4 \sccm\leftskip=0\scpt
plus1fill\rightskip=0\scpt plus1fill\vbox{ \bf
Conformally\vskip-8\scpt Invariant Scalar\vskip-6\scpt Field
$\chi$} }

\setbox11=\hbox{\hsize=5 \sccm\leftskip=0\scpt
plus1fill\rightskip=0\scpt plus1fill\vbox{ {\bf Change of
variable} {\it (redefinition of the scalar)} }}

\setbox12=\hbox{\hsize=5 \sccm\leftskip=0\scpt
plus1fill\rightskip=0\scpt plus1fill\vbox{ {\bf Change of
variables} {\it (arbitrary conformal resc. \piu redefin.~of the
scalar)} }}

 \offinterlineskip \line{\block{6}{3}{\bf Nonlinear Gravity
Theories}{$L=\sqg[\kappa R+aR^2+O(R^3)]$}\freccia \vbox
to3\sccm{\vfill\mblock{5.5}{2}{\bf(A)}{\vbox{\box4}}}
\pfreccia \block{12.3}{3.1}{\bf Scalar--Tensor Gravity Theories}
{\block{6}{2.8}{}{\vbox{\box3}}$\!$
\block{6}{2.8}{}{\vbox{\box1}}}\hfill }\vskip-4\scpt
\line{\vbox to 9.85\sccm{\hbox to1\sccm{\hfill\vrule
height6.15\sccm}}\hskip-1\sccm
\vbox{\mblock{5}{3}{\bf(C=A+B)}{\vbox{\box6}}\vskip0.8
\sccm}\vbox{\hbox to1\sccm {\hrulefill}\vskip30\scpt\hbox
to1\sccm{\hrulefill}\vskip44\scpt}\block {6.1}{3.3}{\bf
Gen.~Relativity + scalar field}{\vbox{\box5}}\vbox{\hbox
to1\sccm {\hrulefill}\vskip30\scpt\hbox
to1\sccm{\hrulefill}\vskip44\scpt}\vbox to 10\sccm{\hbox
to1\sccm {\vrule height3.1\sccm\hfill}\vskip1.9\sccm\hbox
to1\sccm {\vrule height3.45\sccm\hfill}\vfill} \hskip-
4.15\sccm\vbox{\mblock{5}{2}{\bf(B)}{\vbox{\box7}}\vskip5
\sccm}\hskip-0.5\sccm
\vbox{\mblock{5}{2}{\bf(B)}{\vbox{\box8}}\vskip2.5\sccm}
\hskip-0.5\sccm \vbox to 10\sccm{\hbox to1\sccm {\vrule
height5.5\sccm\hfill}\vskip2\sccm\hbox to1\sccm {\vrule
height1.9\sccm\hfill}\vfill}\hskip-8\sccm\vbox{\hbox
to7\sccm {\hrulefill}\vskip8\scpt\hbox
to7.5\sccm{\hrulefill}\vskip8\scpt}%
\vbox{\mblock{5.2}{1.2}{\bf
(D)}{\vbox{\box11}}\vskip0.95\sccm
\hbox{\hskip0.6\sccm\block{4.2}{1.5}{}{\vbox{\box10}}}\vskip
2.38\sccm \mblock{5.2}{2}{\bf (E)}{\vbox{\box9}}} \hskip-
1.5\sccm\vbox to 10\sccm{\hbox to1\sccm {\vrule
height1.5\sccm\hfill}\vskip0.85\sccm {\vrule
height1\sccm\hfill}\vskip1.5\sccm\hbox to1\sccm {\vrule
height3\sccm\hfill}\vfill} \hskip-
35.1\sccm\vbox{\mblock{5.2}{1.65}{\bf(F)}{\vbox{\box12}}
\vskip7.3\sccm} \hskip-5\sccm\vbox{\hbox{\hskip0.7\sccm
\vrule height1\sccm\hskip3\sccm\vrule}\vskip8.85\sccm}
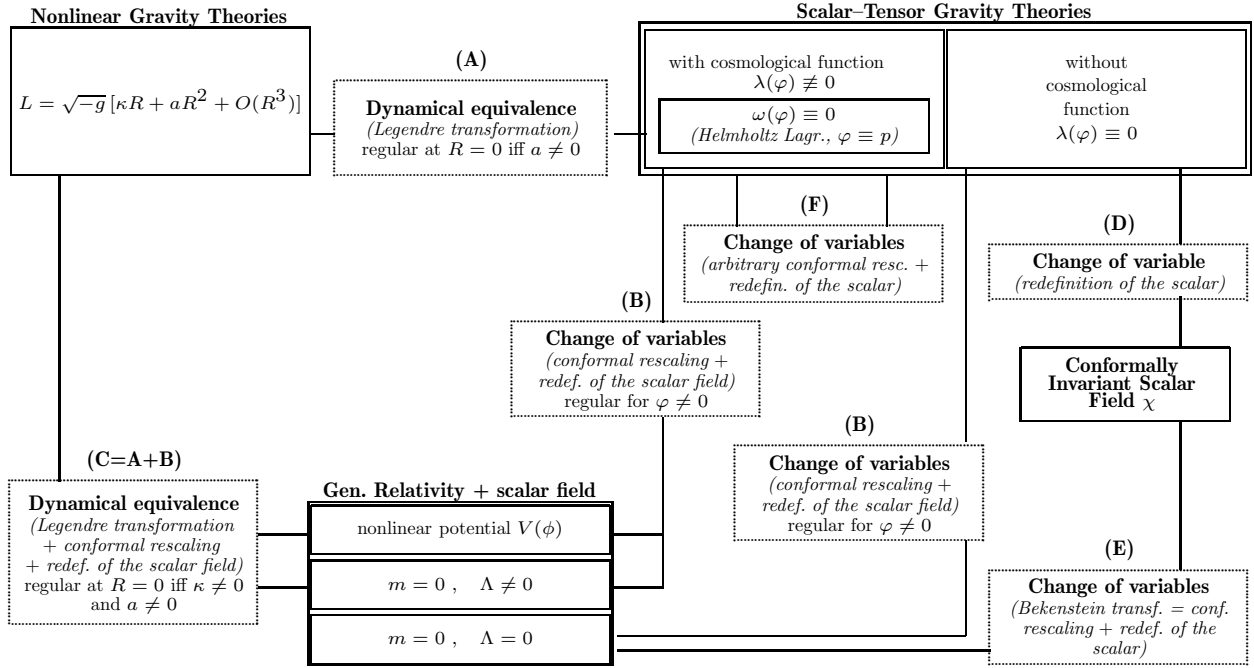
\hfill}} \bigskip {\bf Fig. 1:} Relations among NLG theories (the
Lagrangian (1.1) is expanded in power series around $R=0$),
STG theories and General Relativity. The connection {\bf (A)} is
described in Appendix C,  $(C.1,2)$; {\bf (B)} is represented by
(2.21,22); {\bf (C)} is outlined in Sect.~1 (1.3,6); {\bf (D)} and
{\bf (E)} are discussed at the end of Sect.~2; {\bf (F)} represents
mappings between STG theories with different
$\omega(\varphi)$ (and different $\lambda(\varphi)$, if any),
being a combination of an arbitrary conformal rescaling and a
suitable field redefinition.

\bye